\documentclass[a4paper,11pt]{article}
\usepackage{jheppub} % for details on the use of the package, please see the JINST-author-manual
\usepackage{booktabs} % For better table rules

\usepackage{url}
\usepackage{hyperref}
\usepackage{amsmath}
\usepackage{multirow} % 用于合并行
\usepackage{adjustbox} % 引入adjustbox宏包
\usepackage{setspace} % 引入设置行间距的宏包
\usepackage{array} % 引入array宏包用于设置列对齐方式
\title{\boldmath Prospect for measurement of CP-violating parameters of \(B_s^0 \to \phi\gamma\) at the Tera Z factory}
\author[1,2]{Hengyu Wang}
\author[1,2]{Hanhua Cui}
\author[1,2]{Yongfeng Zhu}
\author[1,2]{Hao Liang}
\author[1,2]{Yuexin Wang}
\author[1,2]{Kaili Zhang}
\author[1,2]{Yi Wang}
\author[1,2]{Weizheng Song}
\author[3]{Lingfeng Li}
\author[1,2]{Yang Zhang}
\author[1,2,4]{Shanzhen Chen*}
\author[1,2,4]{Manqi Ruan*}

\affiliation[1]{Institute of High Energy Physics, Chinese Academy of Sciences\\Beijing 100049, China}
\affiliation[2]{University of Chinese Academy of Sciences(UCAS)\\Beijing 100049, China}
\affiliation[3]{International Center of Theoretical Physics-Asia Pacific,University of Chinese Academy of Sciences(UCAS)\\Beijing 1000049, China}
\affiliation[4]{ Henan Center for High Energy Physics\\Henan 450046, China}

\emailAdd{ruanmq@ihep.ac.cn, szchen@ihep.ac.cn}

\abstract{
The $b\rightarrow s\gamma$ transition is a critical flavor-changing neutral current (FCNC) process that could be used to probe CP parity violation (CPV) and new physics (NP). Future electron-positron colliders operating as Tera-Z factories, such as the Circular Electron-Positron Collider (CEPC) and the Future Circular Collider (FCC-ee), offer unique advantages for studying flavor physics, as they provide large data samples, clean collision environments, and superior detector performance. Based on fast simulation studies, we quantify the anticipated precision for measuring $B_{s}^{0}\rightarrow\phi\gamma$ at the CEPC Z pole operation, showing that the relative statistical uncertainty could be $(0.107 \pm 0.004)\%$, improved by approximately two orders of magnitude compared to existing measurements. Additionally, we perform a time-dependent analysis of the $B_{s}^{0}\rightarrow\phi\gamma$ decay, accounting for $B_{s}^{0}$-$\overline{B}_{s}^{0}$ mixing to extract the mixing-induced and CP-violating parameters $\mathcal{A}_{\phi\gamma}^{\Delta}$, $\mathcal{C}_{\phi\gamma}$, and $\mathcal{S}_{\phi\gamma}$. Using the central values of the LHCb measurements as inputs, the anticipated precisions of these parameters are evaluated as:
\begin{align*}
\boldsymbol{A_{\phi\gamma}^\Delta} &= -0.670 \pm 0.021\text{ (stat.)} \pm 0.040\text{ (syst.)}, \\
\boldsymbol{\mathcal{C}_{\phi\gamma}} &= \phantom{-}0.110 \pm 0.010\text{ (stat.)} \pm 0.008\text{ (syst.)}, \\
\boldsymbol{\mathcal{S}_{\phi\gamma}} &= \phantom{-}0.430 \pm 0.010\text{ (stat.)} \pm 0.010\text{ (syst.)}.
\end{align*}
These promising results not only highlight the physics potential of the CEPC but also provide a critical reference for similar future $e^+e^-$ facilities, notably the FCC-ee.
}

\keywords{Bottom Quarks, Precise Measurement, Flavor-Changing Neutral Currents (FCNC), CP-violating, New Physics}

\begin{document}
\maketitle
\flushbottom

\section{Introduction}
\label{sec:intro}

Flavor-changing neutral current (FCNC) processes serve as a crucial probe of fundamental interactions and  are highly important for testing the Standard Model (SM) \cite{Kumar:2025gsg, Allwicher:2024ncl}. %FCNC transition is suppressed as it occurs at loop level in the SM. %This unique characteristic offers an exceptional experimental platform at the intersection of electroweak-scale physics and quantum chromodynamics (QCD), enabling precision tests of the Standard Model\cite{LHCb:2017avl,LHCb:2023uiv,Halder:2024amk,Paul:2016urs}.
FCNC processes in the SM are suppressed due to the off-diagonal elements of the CKM matrix and loop factors, making such decays rare and difficult to detect \cite{LHCb:2017avl,LHCb:2023uiv,Halder:2024amk,Paul:2016urs}. A significant deviation from SM predictions in FCNC observables often serves as a ``smoking gun" for new physics (NP), which refers to hypothetical particles or interactions not described by the SM. Therefore, a detailed study of FCNC processes can effectively test the SM and potentially uncover NP.

\begin{figure}[h]
  \centering
  \includegraphics[width=0.8\textwidth]{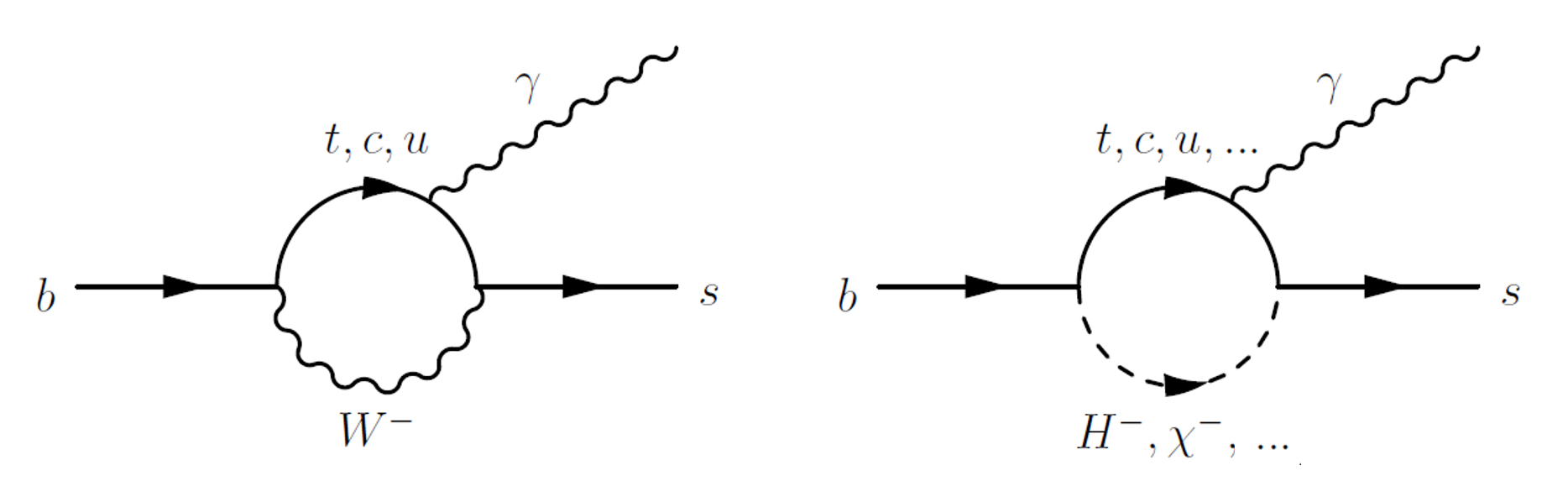}
  \caption{The \(b \to s\gamma\) penguin diagram, mediated by SM particles (left)
and new physics particles (right).}
  \label{fig:penguin_diagram}
\end{figure}

The radiative \(b \to s\gamma\) transition is an FCNC process. This transition occurs in the SM through one-loop \(W\) boson exchange, as shown on the left panel of Figure~\ref{fig:penguin_diagram}. Potential exchange of new virtual particles in the loop may modify the decay rates and the helicity structure of the vertex, as illustrated on the right panel of Figure~\ref{fig:penguin_diagram}. %Significant contributions from right-handed helicity and changes in the decay rates would be clear signatures of new physics, as predicted in several extensions of the Standard Model \cite{Atwood:1997zr}. 
The exclusive decay \(B_s^0 \to \phi\gamma\) \footnote{The inclusion of charge-conjugate processes is implied throughout, unless stated otherwise.} proceeds via the radiative \(b \to s\gamma\) transition, and exhibits a unique feature. Unlike many FCNC processes, its SM-predicted branching fraction is not heavily suppressed, with a theoretical estimate of approximately \(4 \times 10^{-5}\) and an uncertainty of about 30\% \cite{Ball:2006eu,Ali_2008}. This relatively large decay rate has allowed the channel to be experimentally observed by previous experiments \cite{Belle:2007dov,LHCb:2012quo}, thereby opening a unique window for further precision measurements and the introduction of new CP-violating observables at future high-luminosity colliders.

The first experimental observation of this decay was made by the Belle Collaboration in 2008. Based on 23.6 \(\text{fb}^{-1}\) of \(\Upsilon(5S)\) data, the measured branching fraction was \((5.7_{-1.5}^{+1.8} \, _{-1.1}^{+1.2}) \times 10^{-5}\) \cite{Belle:2007dov}. Subsequently, the LHCb Collaboration, using \(1.0\ \text{fb}^{-1}\) of \(\text{pp}\) collision data, measured the ratio of the branching fractions for \(B^0 \to K^{* 0}\gamma\) and \(B_s^0 \to \phi\gamma\) decays and combined it with the world-average value, improving the measurement of \(B_s^0 \to \phi\gamma\) branching fraction to \((3.5 \pm 0.4) \times 10^{-5}\) \cite{LHCb:2012quo}. Belle further analyzed 121.4 \(\text{fb}^{-1}\) of \(\Upsilon(5S)\) data and obtained a more precise result of \((3.6 \pm 0.5 \pm 0.7) \times 10^{-5}\), which is consistent with the theoretical predictions and with the LHCb results \cite{LHCb:2019vks,Belle:2014sac}. Current measurements of this channel by Belle and LHCb have provided valuable insights. However, NP scenarios may hide within current measurement uncertainties. In well-motivated NP frameworks, such as Left-Right Symmetric Models (LRSM) or Supersymmetric (SUSY) models with non-minimal flavor violation, the Wilson coefficient $C_7^\prime$ governing the right-handed photon polarization can be significantly enhanced. This enhancement typically shifts CP-violating observables by $\mathcal{O}(1\% - 10\%)$ \cite{Paul:2016urs,King:2020ldn,PhysRevLett.79.185,Bharucha:2015bzk,Blake:2016olu}. 
Therefore, a substantial improvement in the experimental precision of these observables is strongly motivated. Although the ultimate sensitivity will be limited by theoretical and experimental systematic uncertainties, the large $B_s^0$ sample expected at future Tera-Z factories can significantly reduce the statistical component and provide percent-level constraints on CP-violating observables in a clean $e^+e^-$ environment. This makes the $B_s^0 \to \phi\gamma$ channel a valuable and complementary probe of photon-helicity structure and possible NP effects.
% Furthermore, the Standard Model predictions for these observables currently carry theoretical uncertainties around the sub-percent level. Therefore, achieving sub-percent level precision is the essential experimental threshold to unambiguously disentangle potential NP effects from the SM background, whereas sub-permille precision would be largely obscured by irreducible hadronic theoretical uncertainties.
%However, new physics scenarios such as supersymmetric models(SUSY) may still contribute to this decay by modifying the dipole operators \cite{Gemintern:2004bw,Grossman:1996era}. Therefore, the precise measurement of this branching fraction is crucial for testing the Standard Model and exploring for new physics.

\begin{table}[h]
    \centering
    \caption{Expected yields of b-hadrons at Belle II, LHCb Upgrade II, and CEPC Tera-Z.}
    \label{tab:b_hadron_yields}
    \begin{tabular}{|c|c|c|c|}
    \hline
    b-hadrons & Belle II & LHCb $(300\ fb^{-1})$ & CEPC$(100\ ab^{-1})$ \\
    \hline
    $B^0, \bar{B}^0$ & $5.4\times 10^{10}\ (50\ ab^{-1}\ on\ \Upsilon(4S))$ & $3\times 10^{13}$ & $4.8\times 10^{11}$ \\

    $B^{\pm}$ & $5.7\times 10^{10}\ (50\ ab^{-1}\ on\ \Upsilon(4S))$ & $3\times 10^{13}$ & $4.8\times 10^{11}$ \\

    $B_s^0, \bar{B}_s^0$ &$ 6.0\times 10^{8}\ (5\ ab^{-1}\ on\ \Upsilon(5S))$ & $1\times 10^{13}$ & $1.2\times 10^{11}$ \\

    $B_c^{\pm}$ & - & $1\times 10^{11}$ & $7.4\times 10^{8}$ \\

    $\Lambda_b^0, \bar{\Lambda}_b^0$ & - & $2\times 10^{13}$ & $1.0\times 10^{11}$ \\
    \hline
    \end{tabular}
\end{table}

The primary physics objective of the CEPC is to identify new physics, especially via the Higgs portal. Furthermore, this future collider could be sensitive to new physics through the flavor physics measurements \cite{Ai:2024nmn}. The facility is expected to operate for over a decade, including two years at the Z pole. Under an operation scenario of 50 MW synchrotron radiation (SR) power per beam \cite{CEPCStudyGroup:2023quu}, it is expected to produce $4.1 \times 10^{12}$ Z bosons over this two-year Z-pole operation period \cite{CEPCStudyGroup:2023quu}, providing a solid foundation for the precise measurement of rare decays. Thanks to the substantial $Z \to b\bar{b}$ branching fraction, this massive Z boson sample translates into an unprecedented number of b-hadrons. Table~\ref{tab:b_hadron_yields} summarizes the expected numbers of b hadrons produced at Belle II \cite{BelleII:2018jsg}, LHCb Upgrade II \cite{LHCb:2018roe,LHCb:2016qpe,LHCb:2019fns,LHCb:2019tea}, and CEPC Tera-Z \cite{HFLAV:2019otj,Zheng:2019gnb}. The production yields of $B^{0}/\bar{B}^{0}$ and \(B^{\pm}\) at CEPC are \(4.8 \times 10^{11}\), approximately an order of magnitude larger than the corresponding yields at Belle II. Furthermore, the production yields of $B^{0}/\bar{B}^{0}$ are \(1.3 \times 10^{11}\), nearly 200 times those of Belle II. For heavier hadrons, such as \(B_{c}^{\pm}\) and $\Lambda_{b}/\bar{\Lambda}_{b}$, CEPC also demonstrates production advantages, with yields of \(7.4 \times 10^{8}\) and \(1.0 \times 10^{11}\), respectively. These abundant hadronic yields will allow the CEPC to accumulate significantly larger data samples for the study of rare b decays \cite{Li:2020bvr,Wang:2022nrm,Li:2022tov,Zheng:2020ult}, substantially enhancing the accuracy and reliability of measurements. Compared to the hadronic collision environment at LHCb, the $e^+e^-$ collisions at the CEPC provide a well-defined initial state with negligible quantum chromodynamics (QCD) underlying events and minimal pile-up effects \cite{Che:2022dig}. This pristine environment significantly suppresses combinatorial backgrounds and further improves the overall measurement precision.

\begin{figure}[htbp]
\centering
\includegraphics[width=.75\textwidth]{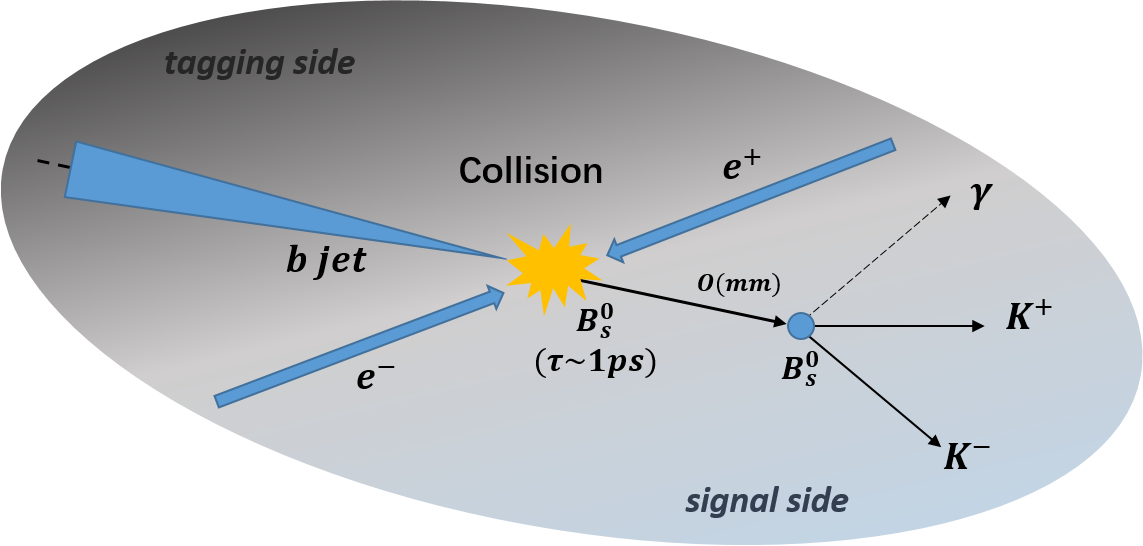}
\caption{The topology of \(B_s^0 \to \phi\gamma\) decay at Z-pole.\label{fig:topology}}
\end{figure}

Figure~\ref{fig:topology} illustrates the decay topology of \(B_{s}^{0} \to \phi(K^{+}K^{-})\gamma\) at the CEPC Z pole. In this scenario, an event is divided into a tagging side and a signal side. The Jet origin identification (JOI) algorithm \cite{Liang:2023wpt} is applied for $b$-jet tagging on the tagging side. \(B_{s}^{0}\) mesons are produced through the hadronization of a $b$ quark originating from the interaction point (IP) on the signal side. The \(B_{s}^{0}\) meson decays into a \(\phi\) meson and a \(\gamma\) photon. The produced \(\phi\) meson subsequently decays into two charged \(K\) mesons, i.e., \(\phi \to K^{+}K^{-}\). The lifetime of the \(B_{s}^{0}\) meson is \((1.516\pm0.006)\times 10^{-12}s\) \cite{ParticleDataGroup:2024cfk}. Given an average momentum of approximately 45 $\mathrm{GeV}/c$, the corresponding flight distance is on the order of \(O(mm)\).

The signal efficiency and purity are defined as follows:
\begin{align}
\epsilon=\frac{N_{\text{rec, corr}}}{N_{\text{all}}},\label{eq:1}\\
p=\frac{N_{\text{rec, corr}}}{N_{\text{rec, total}}}, \label{eq:2}
\end{align}
where $N_{\text{rec, corr}}$ is the number of correctly reconstructed $B_{s}^{0} \to \phi\gamma$ events, $N_{\text{all}}$ is the total number of generated signal events, and $N_{\text{rec, total}}$ is the total number of reconstructed signal candidates. The estimated statistical uncertainty for the reconstructed $B_{s}^{0} \to \phi\gamma$ signal candidates is defined as $\frac{\sqrt{S + B}}{S}$, which is proportional to $\frac{1}{\sqrt{\epsilon\times p}}$, where $S$ and $B$ denote the expected number of signal and background events, respectively. In this analysis, the event selection criteria are optimized to maximize the measurement precision.

This paper utilizes fast simulation, a cut-based method, and boosted decision trees (BDT) \cite{Coadou:2022nsh} to evaluate the anticipated measurement precision of this decay channel and to estimate the expected uncertainties of the CP-violating parameters at the CEPC. The remainder of this paper is organized as follows: Section 2 introduces the reference detector, samples, and the fast simulation method. Section 3 presents the detailed event selection strategy, and results from BDT training. Section 4 presents the projected sensitivities of the CP-violating parameters derived from a time-dependent analysis of the $B_s^0 \to \phi\gamma$ decay. Finally, conclusions are drawn in Section 5.

\section{Reference detector, samples, and method}

\subsection{CEPC baseline detector}

\begin{figure}[htbp]
\centering
\includegraphics[width=.8\textwidth]{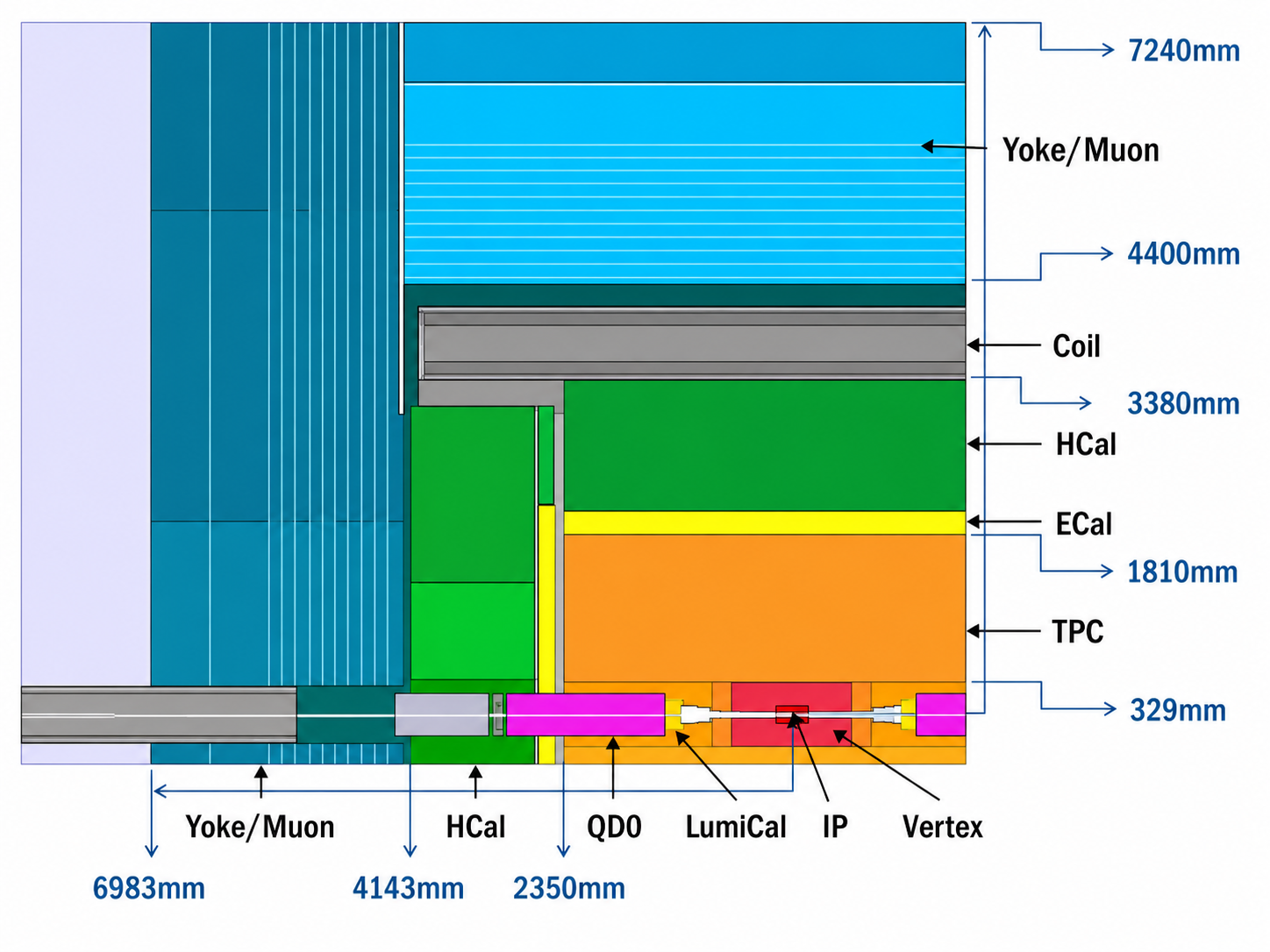}
\caption{Schematic cross-sectional layout of the baseline CEPC detector design. From inner to outer, the detector is composed of a silicon pixel vertex detector, a silicon-strip inner tracker, a TPC, a silicon-strip external tracker, an ECAL, a HCAL, a 3~T solenoid, and a return yoke instrumented with a muon detector.\label{fig:CEPCCDR}}
\end{figure}

The CEPC CDR baseline detector schematic \cite{CEPCStudyGroup:2018ghi} is shown in Figure~\ref{fig:CEPCCDR}. The innermost sub-detector is the tracking system, designed to reconstruct charged particles and measure their momenta. It comprises a barrel region and an endcap region. The barrel region includes a three-layer silicon pixel vertex detector (VTX), a two-layer silicon-strip inner tracker (SIT), a Time Projection Chamber (TPC), and a single-layer silicon-strip external tracker (SET). The endcap region consists of five pairs of silicon tracking disks forming the Forward Tracking Detector (FTD), located on both sides of the interaction point (IP). The two innermost disk pairs employ pixel sensors to cope with the high hit density close to the beam pipe, while the three outer disk pairs employ microstrip sensors. According to the CEPC CDR \cite{CEPCStudyGroup:2018ghi}, the total expected material budget of the tracking detectors upstream of the electromagnetic calorimeter (ECAL) is approximately \(0.10\,X_0\) in the barrel region and \(0.40\,X_0\) in the endcap region. The hadronic calorimeter (HCAL) is located outside the ECAL. The transverse impact-parameter and transverse-momentum resolutions used in this study are parameterized as follows~\cite{CEPCStudyGroup:2018ghi}:
\begin{align}
\sigma_{r\phi} &= 5 \oplus \frac{10}{p \sin^{3/2}\theta}\ (\mu\text{m}), \label{eq:VTX}\\
\sigma(1/p_T) &= 2 \times 10^{-5} \oplus \frac{0.001}{p \sin^{3/2}\theta}\ (\left(\mathrm{GeV}/c\right)^{-1}), \label{eq:Momentum}
\end{align}
where $p$ and $p_T$ denote the track momentum and transverse momentum in GeV/$c$, respectively, and $\theta$ is the polar angle with respect to the beam direction. In the standard CEPC coordinate system, the $z$-axis is aligned with the beamline, while $r$ and $\phi$ denote the radial distance and the azimuthal angle in the transverse plane, respectively. Accordingly, Equation~\ref{eq:VTX} describes the transverse impact parameter resolution at the track origin in the $r-\phi$ plane, and Equation~\ref{eq:Momentum} describes the track transverse momentum measurement resolution.

Surrounding the tracking system is the calorimeter system. A silicon-tungsten sampling electromagnetic calorimeter (Si-W ECAL) measures electromagnetic showers, while a steel-glass resistive plate chambers sampling HCAL measures hadronic energy deposits. Based on the CEPC Conceptual Design Report (CDR) \cite{CEPCStudyGroup:2018ghi,Kawagoe:2019dzh}, the Si-W ECAL is confirmed as the baseline option. Its performance is parameterized with an energy resolution of: 
\begin{align}
\frac{\sigma_E}{E} &= \frac{17\%}{\sqrt{E}} \oplus 1\%, \label{eq:ECAL}
\end{align}
where $E$ is the $\gamma$ energy in GeV. In addition to energy measurements, the Si-W ECAL provides time-of-flight (TOF) information with a precise time resolution of 50 ps, which significantly enhances the particle identification (PID) capabilities \cite{Zhu:2022hyy}. 
%This study relies firmly on the CDR baseline setup. 
%it is important to note that recent R$\&$D campaigns for the CEPC Reference Detector Technical Design Report (TDR) have proposed novel high-resolution crystal calorimeter designs. The designed energy resolution of better than $3.0\%/\sqrt{E} \oplus 1\%$ \cite{CEPCStudyGroup:2025kmw}. To maintain a well-defined reference, our primary simulation strictly adopts the CDR Si-W parameters. However, the potential sensitivity improvements brought by these updated TDR ECAL configurations are specifically discussed and evaluated later in Section 4.  

\begin{figure}[h]
\centering
\includegraphics[width=1.\textwidth]{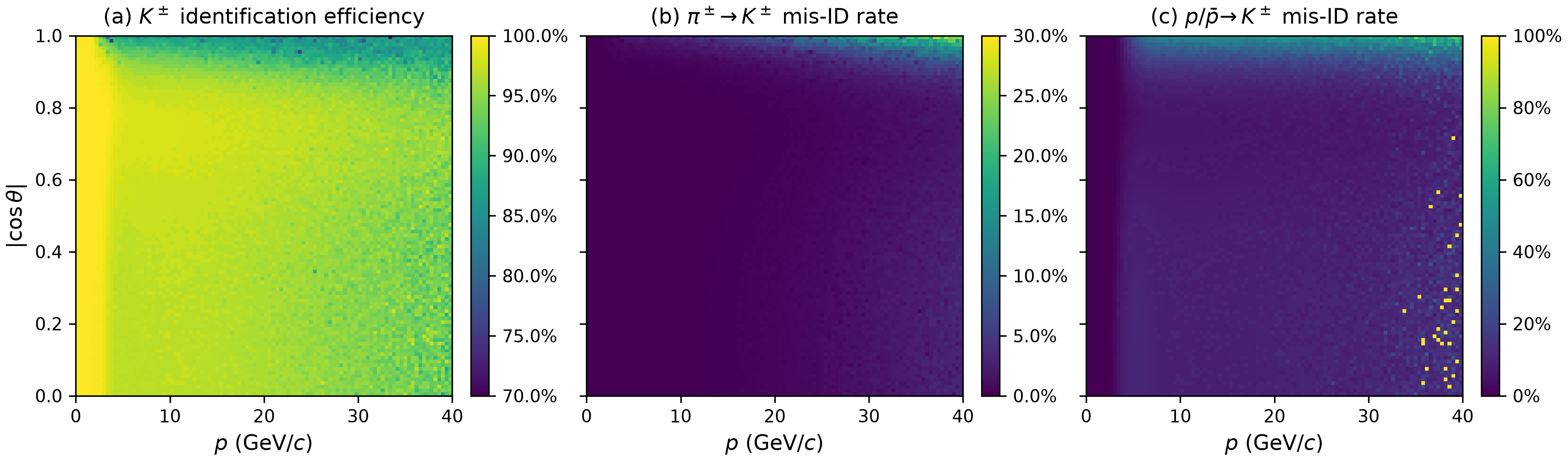}

\caption{Two-dimensional particle-identification performance obtained by
combining \(dE/dx\) and TOF information~\cite{Zhu:2022hyy}:
(a) the \(K^\pm\) identification efficiency,
(b) the \(\pi^\pm\to K^\pm\) misidentification rate, and
(c) the \(p/\bar{p}\to K^\pm\) misidentification rate,
shown as functions of \(p\) and \(|\cos\theta|\).
The color scale in each panel represents the corresponding efficiency
or misidentification rate.} \label{fig:PIDmatrix}
\end{figure}

The PID performance is important for suppressing backgrounds with charged pion or proton misidentification in the reconstruction of \(\phi \to K^+K^-\) candidates. In the CEPC baseline design, the PID capability is primarily provided by combining the specific energy loss ($dE/dx$) measured by the TPC and the precise TOF measurements. The PID performance depends strongly on both the track momentum and the polar angle. In this study, the momentum- and angular-dependent \(K^\pm\) identification efficiency and misidentification rate are taken from the two-dimensional PID performance matrices shown in Figure~\ref{fig:PIDmatrix}~\cite{Zhu:2022hyy}. These matrices are used in the fast simulation to model the realistic PID response across different track momentum and polar-angle bins.

\subsection{Simulation setup}

This study employs the Marlin fast simulation package \cite{Gaede:2006pj} within the CEPC software framework. The fast simulation models the detector response by smearing the kinematic properties of generated particles based on the baseline CEPC detector performances described in Section 2.1. 

For charged tracks, the momenta and impact parameters of charged particles are smeared according to the parameterized tracking resolutions defined in Equations~\ref{eq:VTX} and \ref{eq:Momentum}. Similarly, the energies of photons are smeared using the energy resolution of the baseline Si-W ECAL, as formulated in Equation~\ref{eq:ECAL}. 
% It is important to emphasize that to maintain a consistent baseline reference and avoid ambiguity, all primary simulated samples in this section are generated strictly using this Si-W ECAL configuration.

Furthermore, our fast simulation adopts a parameterized approach for PID. To accurately reflect realistic experimental conditions, we do not assume an ideal, flat $3\sigma$ separation power across the entire kinematic range. Instead, the background suppression algorithm evaluates the true identity of each track and assigns a reconstructed PID based on the realistic multidimensional identification-efficiency and misidentification-rate matrices shown in Figure~\ref{fig:PIDmatrix}. This ensures that the simulated PID performance accurately integrates the momentum and angular dependence of the CEPC combined tracking and TOF systems.

\subsection{Samples}

The signal events for \( B_s^0 \to \phi \gamma \) primarily originate from the hadronization process \( b \to B_s^0 \). The expected yields depend on the relevant branching fractions and fragmentation fractions, which are summarized in Table~\ref{tab:Br_of_Bs2phigamma}. According to existing experimental data, the branching fraction for the $Z$ boson decaying into bottom quark pairs is \(\mathcal{B}(Z \to b\bar{b}) = (15.12 \pm 0.05)\% \) \cite{ParticleDataGroup:2024cfk}. For the hadronization process, it is well known that the $b$-quark fragmentation fractions depend on the center-of-mass energy. Since the CEPC will operate exactly at the $Z$ pole, we can directly adopt the fragmentation fractions measured in $Z$ decays, where the fraction for \( b \to B_s^0 \) is \(f(b \to B_s^0) = 0.101 \pm 0.008 \) \cite{HFLAV:2019otj}. Furthermore, the branching fraction \( \mathcal{B}(B_s^0 \to \phi \gamma) \) is \((3.4 \pm 0.4) \times 10^{-5} \) \cite{ParticleDataGroup:2024cfk}. In this study, we reconstruct the \( \phi \) meson using the \( \phi \to K^+ K^- \) decay mode, which has a branching fraction of \(\mathcal{B}(\phi \to K^+ K^-) = (49.1 \pm 0.5)\% \) \cite{ParticleDataGroup:2024cfk}.

\begin{table}[h]
    \centering
    \caption{Branching fractions and fragmentation fractions for \( B_s^0 \to \phi \gamma \) decays.}
    \label{tab:Br_of_Bs2phigamma}
    \begin{tabular}{lcc}
    \toprule
    Parameter & Value & Source \\
    \midrule
        \( \mathcal{B}(Z \to b\bar{b}) \) & (\(15.12 \pm 0.05\))\% & PDG \cite{ParticleDataGroup:2024cfk} \\ 
        \( f(b \to B^0) \) & \( 0.404 \pm 0.007 \) & HFLAV \cite{HFLAV:2019otj} \\ 
        \( f(b \to B_s^0) \) & \( 0.101 \pm 0.008 \) & HFLAV \cite{HFLAV:2019otj} \\ 
        \( \mathcal{B}(B_s^0 \to \phi \gamma) \) & \( (3.4 \pm 0.4) \times 10^{-5} \) & PDG \cite{ParticleDataGroup:2024cfk} \\ 
        \( \mathcal{B}(\phi \to K^+ K^-) \) & \( (49.1 \pm 0.5)\)\% & PDG \cite{ParticleDataGroup:2024cfk} \\
    \bottomrule
    \end{tabular}
\end{table}

Based on the expected CEPC Tera-$Z$ yield of $N_Z = 4.1 \times 10^{12}$ $Z$ bosons, and including the charge-conjugate state, the expected number of signal events for the decay \( B_s^0 \to \phi (K^+ K^-) \gamma \), denoted as $N_{\text{sig}}$, can be estimated as follows:
\begin{align}
N_{\text{sig}} &= \mathbf{2} \times N_Z \times \mathcal{B}(Z \to b\bar{b}) \times f(b \to B_s^0) \times \mathcal{B}(B_s^0 \to \phi \gamma) \times \mathcal{B}(\phi \to K^+ K^-) \nonumber \\
&\approx 2.1 \times 10^6. \label{eq:yield}
\end{align}

Table~\ref{tab:teraZ_yields} summarizes the expected yields, the simulated sample sizes, and the event weights of the signal and inclusive \( Z \to q\bar{q} \) backgrounds used in this analysis. The background was generated by Whizard \cite{Kilian:2007gr} and the signal samples was generated by Pythia \cite{Sjostrand:2014zea} packages. Given the extremely large expected yield at the Tera-$Z$, producing the full dataset of background events is computationally prohibitive. Consequently, to balance computational feasibility with the need for robust background modeling, this study utilizes a generated sample of \( 10^9 \) inclusive \( Z \to q\bar{q} \) events, scaled by their respective event weights. These samples were generated with the \(B_s^0 \to \phi \gamma\) signal removed to avoid double-counting, and are categorized according to the quark-antiquark flavor (\(d\bar{d}, u\bar{u}, s\bar{s}, c\bar{c}, b\bar{b}\)).
\begin{table}[htbp]
    \centering
    \caption{Expected yields, simulated sample sizes, and the corresponding event weights for the signal and backgrounds at the CEPC Tera-$Z$. The weight is defined as the ratio of the expected yield to the simulated sample size.}
    \label{tab:teraZ_yields}
    \begin{tabular}{lccc}
        \toprule
        Process & Expected Yield & Simulated Samples & Weight \\ 
        \midrule
        \( B_s^0 \to \phi(K^+K^-) \gamma \) & \( 2.1 \times 10^6 \) & \( 2.2 \times 10^5 \) & \(9.5 \) \\ 
        \(Z \to d\bar{d}\) & \( 6.5 \times 10^{11} \) & \( 2.0 \times 10^8 \) & \( 3250 \) \\ 
        \(Z \to u\bar{u}\) & \( 4.6 \times 10^{11} \) & \( 2.0 \times 10^8 \) & \( 2300 \) \\ 
        \(Z \to s\bar{s}\) & \( 6.5 \times 10^{11} \) & \( 2.0 \times 10^8 \) & \( 3250 \) \\ 
        \(Z \to c\bar{c}\) & \( 4.9 \times 10^{11} \) & \( 2.0 \times 10^8 \) & \( 2450 \) \\ 
        \(Z \to b\bar{b}\) & \( 6.2 \times 10^{11} \) & \( 2.0 \times 10^8 \) & \( 3100 \) \\ 
        \bottomrule
    \end{tabular}
\end{table}

Although Bhabha scattering and other leptonic $Z$ decays possess enormous cross-sections at the CEPC Tera-$Z$ operation, these events can be easily vetoed thanks to the excellent lepton identification and distinct track multiplicities \cite{Yu:2017mpx}. Consequently, the inclusive $Z \to q\bar{q}$ sample serves as the primary background source in this analysis.

Among the hadronic $Z$ decays, the background sources are broadly classified into random combinatorial backgrounds, partially reconstructed backgrounds, and peaking backgrounds. The random combinatorial background primarily arises from inclusive $Z \to b\bar{b}$, $Z \to c\bar{c}$, and $Z \to s\bar{s}$ processes, where random track and photon combinations from fragmentation and displaced vertices can mimic the signal topology. Charm-related contributions from $Z \to c\bar{c}$ events are therefore included in this category.

Partially reconstructed backgrounds arise when one or more final-state particles escape reconstruction. For example, decays such as $B^+ \to \phi(K^+K^-) K^+ \gamma$ can create structures in the lower mass sideband when a kaon escapes reconstruction. Secondary charm produced in $b$-hadron decays in $Z \to b\bar{b}$ events can also contribute to this category, since charm-hadron decays may contain real $K^+K^-$ or $\phi$ candidates that are combined with photons from fragmentation or neutral-meson decays.
Peaking backgrounds arise when particles are misidentified in exclusive heavy-flavor radiative decays with topologies close to the signal. Examples include $\Lambda_b \to \Lambda^*(pK^-) \gamma$ and $B^0 \to K^{*0}(K^+\pi^-) \gamma$, where a proton or pion is incorrectly identified as a kaon. Additionally, decays involving a high-energy $\pi^0$, such as $B^0 \to \phi(K^+K^-)\pi^0(\gamma\gamma)$, can fake the prompt signal photon if the two decay photons merge or one escapes reconstruction. These contributions are expected to be suppressed to the 0.1\% level through PID and strict $\phi$ mass constraints. By generating $10^9$ weighted inclusive $Z \to q\bar{q}$ events and applying the parameterized multi-dimensional PID matrices and ECAL resolutions, our fast simulation includes the large combinatorial phase space, charm-related contributions, and the realistic misidentification rates of these physical backgrounds.

Despite capturing the primary background features, we acknowledge the inherent limitations of our simulation framework. At the generator level, the kinematics and multiplicity of the combinatorial background depend strongly on the phenomenological QCD parton shower and hadronization models in Whizard \cite{Kilian:2007gr} and Pythia \cite{Sjostrand:2014zea}. At the detector level, our fast simulation might not fully encapsulate complex effects, including non-Gaussian resolution tails, secondary material interactions, and beam-induced backgrounds. Consequently, we explicitly acknowledge that the exact background composition and the final projected sensitivities will likely change significantly once a fully realistic detector is introduced.

\section{Events selection and results}

\subsection{Events selection}

In the event selection process, the kinematic features and reconstructed masses of $\phi$ and $B_s^0$ candidates are used to distinguish the signal from backgrounds. The selection criteria are optimized iteratively. For each discriminating variable, the optimal cut value is determined by minimizing the relative statistical uncertainty, defined as $\sqrt{S+B}/S$, where $S$ and $B$ are the expected signal and background yields, respectively. Subsequently, $b$-tagging is applied to further suppress the remaining backgrounds. The detailed selection steps are outlined below.

\subsubsection{Reconstruction of $\phi\to K^+K^-$}

To accurately identify \(B_{s}^{0} \to \phi \gamma\) signal events, we first reconstruct the $\phi$ meson via its decay \(\phi \to K^+K^-\). The CEPC's tracking system enables vertex reconstruction with a precision better than 5~$\mu$m. Vertex reconstruction is performed using the LCFIPlus package \cite{LCFIPlus}. This capability is crucial for suppressing the QCD combinatorial background originating directly from the nominal beam collision point. 

To separate the signal from the combinatorial background, the selection of $\phi$ candidates is performed in four sequential steps. The exact process proceeds as follows:

First, as a baseline requirement, events containing oppositely charged track pairs identified by the PID system as kaons ($K^\pm$) are selected to form the initial $\phi$ candidates. 

\begin{figure}[h]
\centering
\includegraphics[width=.45\textwidth]{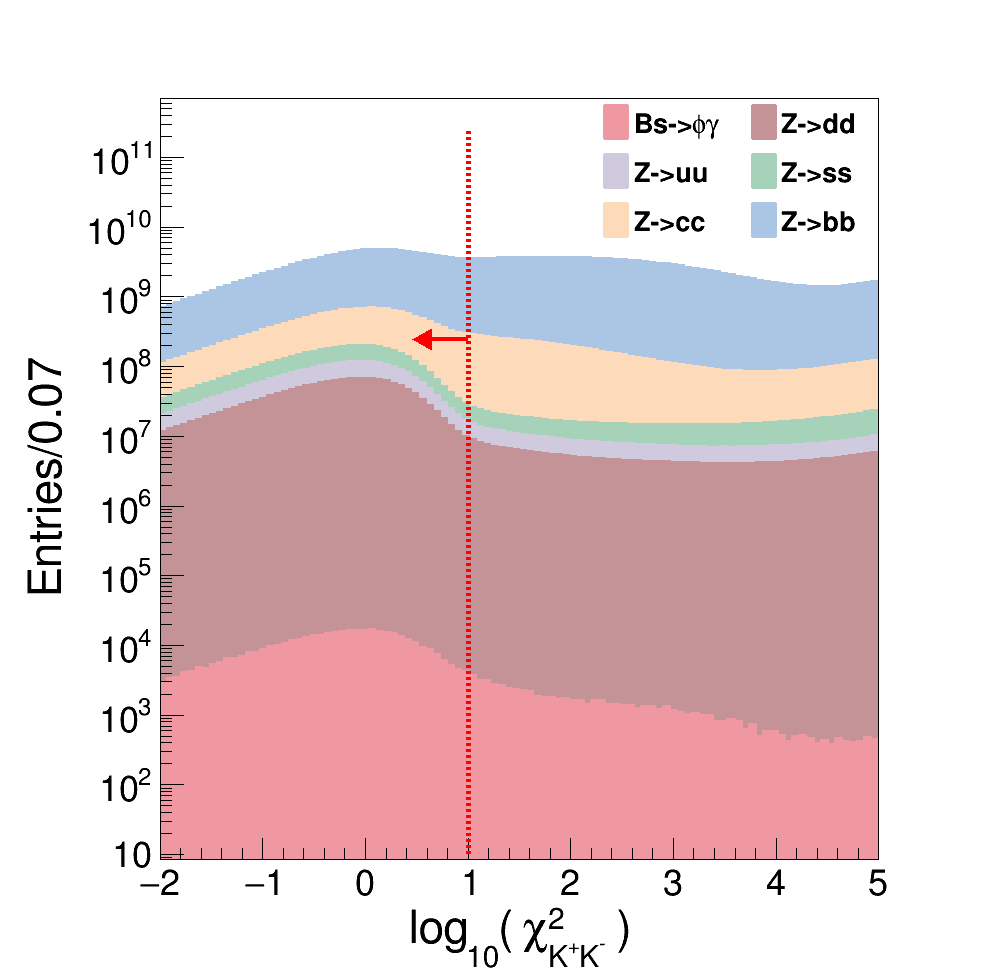}
\includegraphics[width=.45\textwidth]{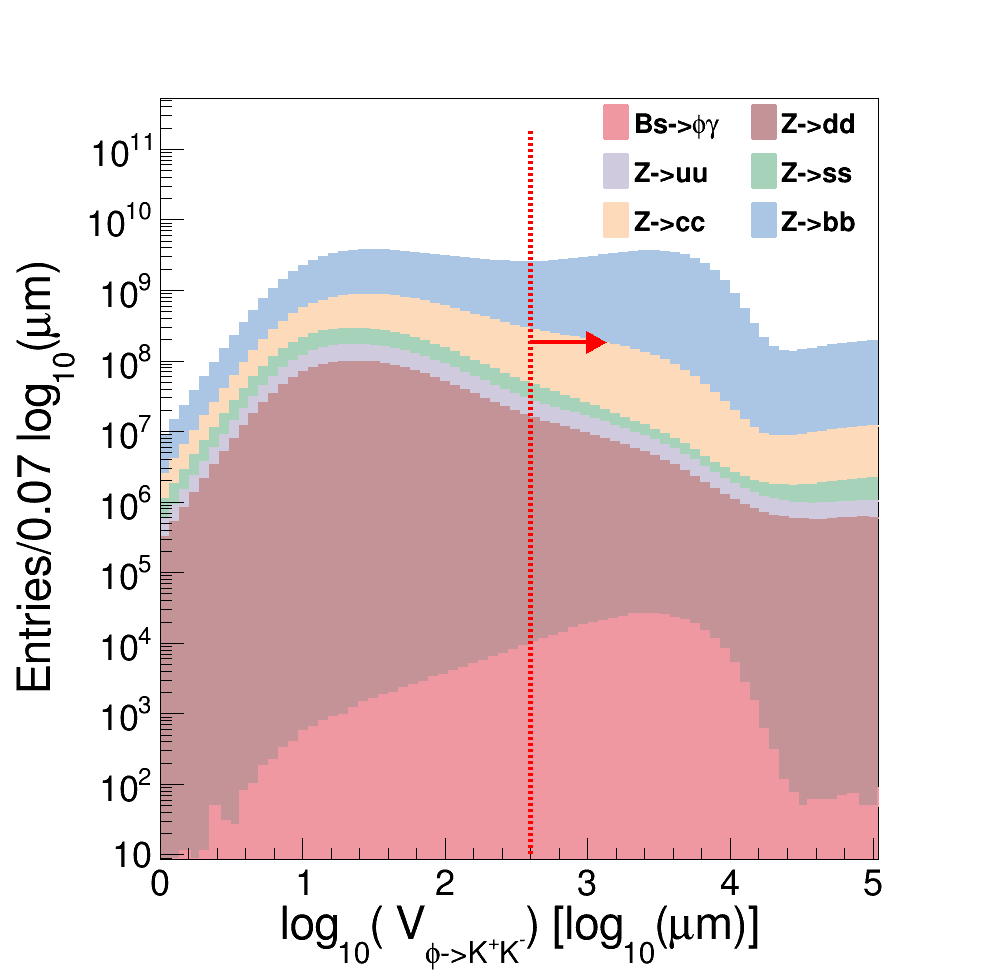}
%\includegraphics[width=.32\textwidth]{image/EventSelection/3PhiMassCut.png}
% \includegraphics[width=.32\textwidth]{image/EventSelection/1Accuracy_vs_Chi2Cut.png}
% \includegraphics[width=.32\textwidth]{image/EventSelection/2Accuracy_vs_VtxDisCut.png}
% \includegraphics[width=.32\textwidth]{image/EventSelection/3Accuracy_vs_PhiMassCut.png}

%\subcaptionbox{Without SET}{\includegraphics[width=.23\textwidth]{image/Distribution_WithOutSET/Distribution_WithOutSET_AngleGaussian_50GeV_Smear50GeV_页面_07.png}}

\caption{Distributions of the vertex-fit quality variable \(\chi^2\) (left) and the \(\phi\)-candidate vertex displacement \(V_{\phi\to K^+K^-}\) (right) for signal and background events. The variable \(V_{\phi\to K^+K^-}\) is defined as the distance between the reconstructed \(K^+K^-\) decay vertex and the interaction point. The selected ranges are \(\log_{10}(\chi^2)<1\) and \(\log_{10}(V_{\phi\to K^+K^-}/\mu\mathrm{m})>2.6\).\label{fig:CutChain_Step2}}
\end{figure}

Next, we evaluate the quality of the vertex fit using the normalized vertex fit parameter, $\chi^2$. This variable is defined as $\chi^2 = \sum_{i=1}^{2} \left( \frac{|V_i - V_{\text{fit}}|}{\sigma_i} \right)^2$, where $V_{\text{fit}}$ is the fitted vertex position of the $\phi$ decay, $V_i$ is the point of closest approach on the $i$-th track to the fitted vertex, and $\sigma_i$ is the corresponding position uncertainty. As shown in the left panel of Figure~\ref{fig:CutChain_Step2}, the $\chi^2$ distribution for the correctly reconstructed signal candidates is noticeably concentrated at lower values compared to the combinatorial background. By minimizing $\sqrt{S+B}/S$, a selection of $\log_{10}(\chi^2) < 1$ is applied to ensure high-quality vertex fits.

Furthermore, we consider the decay vertex distance of the $\phi$ candidate, denoted as $V_{\phi\to K^+K^-}$. Because the $B_s^0$ meson travels a macroscopic distance of $\mathcal{O}(\text{mm})$ before decaying, the vertex of $\phi$ meson in the signal process is significantly displaced from the interaction point (IP), whereas the combinatorial background largely originates directly from the IP. The right panel of Figure~\ref{fig:CutChain_Step2} illustrates this distinct topological difference. We minimize $\sqrt{S+B}/S$ and apply a displaced vertex requirement of $\log_{10}(V_{\phi\to K^+K^-}/\mu\text{m}) > 2.6$.

\begin{figure}[h]
    \centering
    \includegraphics[width=.45\textwidth]{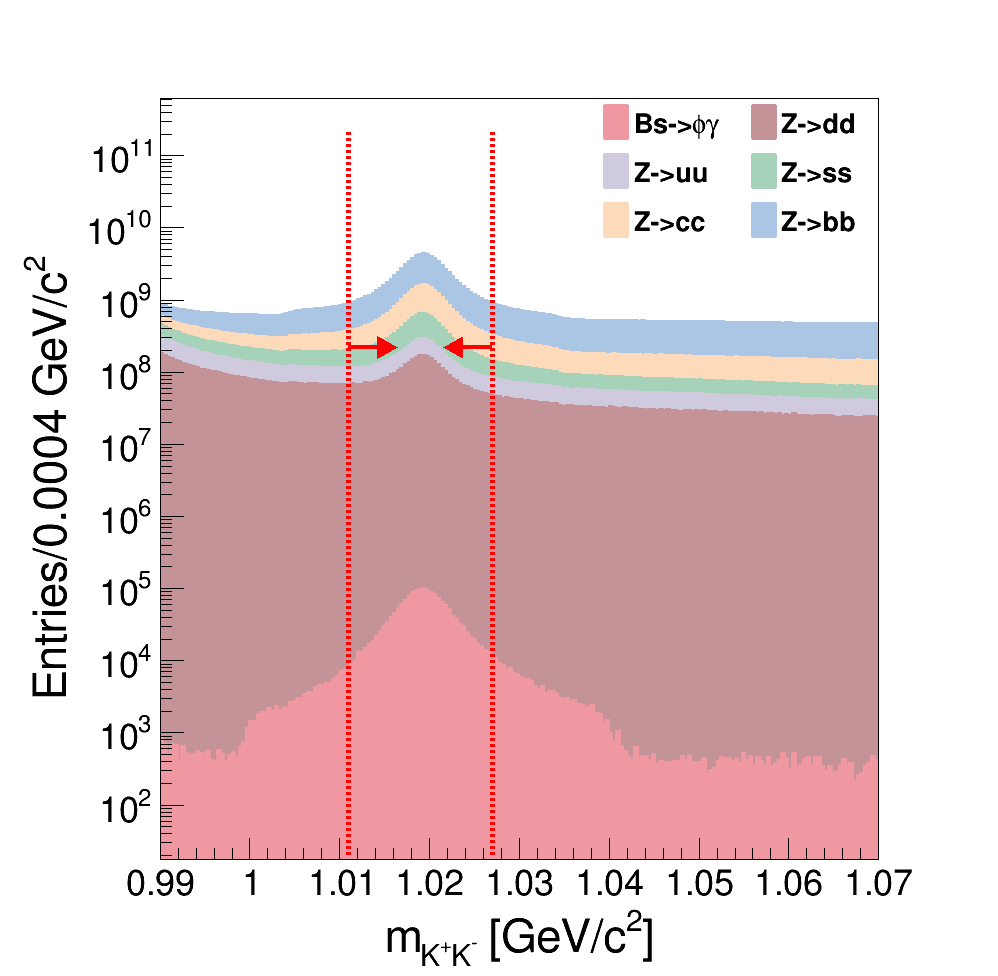}
    \caption{The invariant mass distribution of the reconstructed $\phi$ candidates. The vertical axis represents the normalized number of events. The selected range is $1.011 < m_{K^+K^-} < 1.027$~$\mathrm{GeV}/c^2$.\label{fig:phi mass}}
\end{figure}

Finally, we examine the invariant mass of the reconstructed $K^+ K^-$ pair, denoted as $m_{K^+ K^-}$. The signal of $\phi$ meson appears as a narrow resonance peak, as displayed in the mass distribution in Figure~\ref{fig:phi mass}. A selection of $1.011 < m_{K^+K^-} < 1.027$~$\mathrm{GeV}/c^2$ is applied based on the optimization of $\sqrt{S+B}/S$.

\subsubsection{Reconstruction of $B_s^0\to\phi\gamma$}

Following the $\phi$ meson selection, we proceed to reconstruct the $B_s^0$ candidates. Because the $B_s^0$ mesons originating from $Z \to b\bar{b}$ hadronization are produced with a significant boost, their decay products are highly collimated and exhibit relatively high momentum. Therefore, we pair the selected $\phi$ candidate with a $\gamma$ to form a $B_s^0$ candidate and introduce two sequential kinematic selection steps based on this distinct topology.

\begin{figure}[h]
\centering
\includegraphics[width=.45\textwidth]{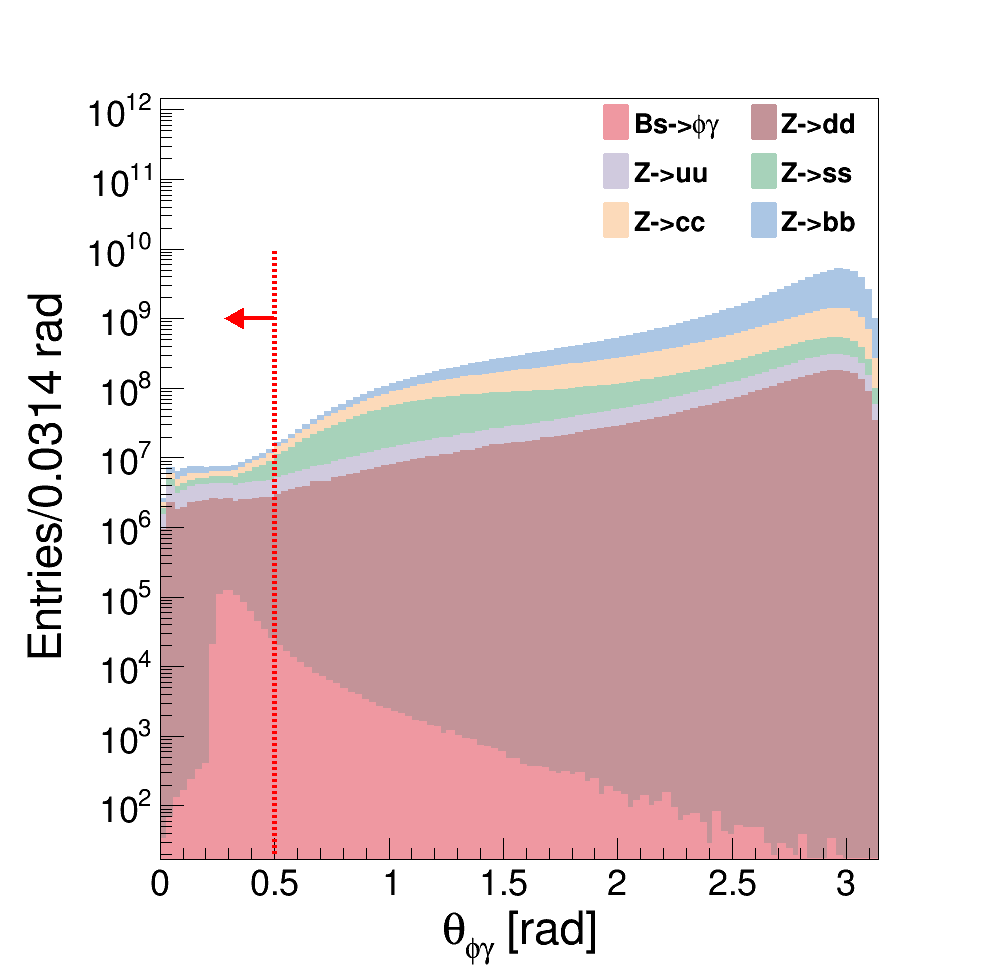}
\includegraphics[width=.45\textwidth]{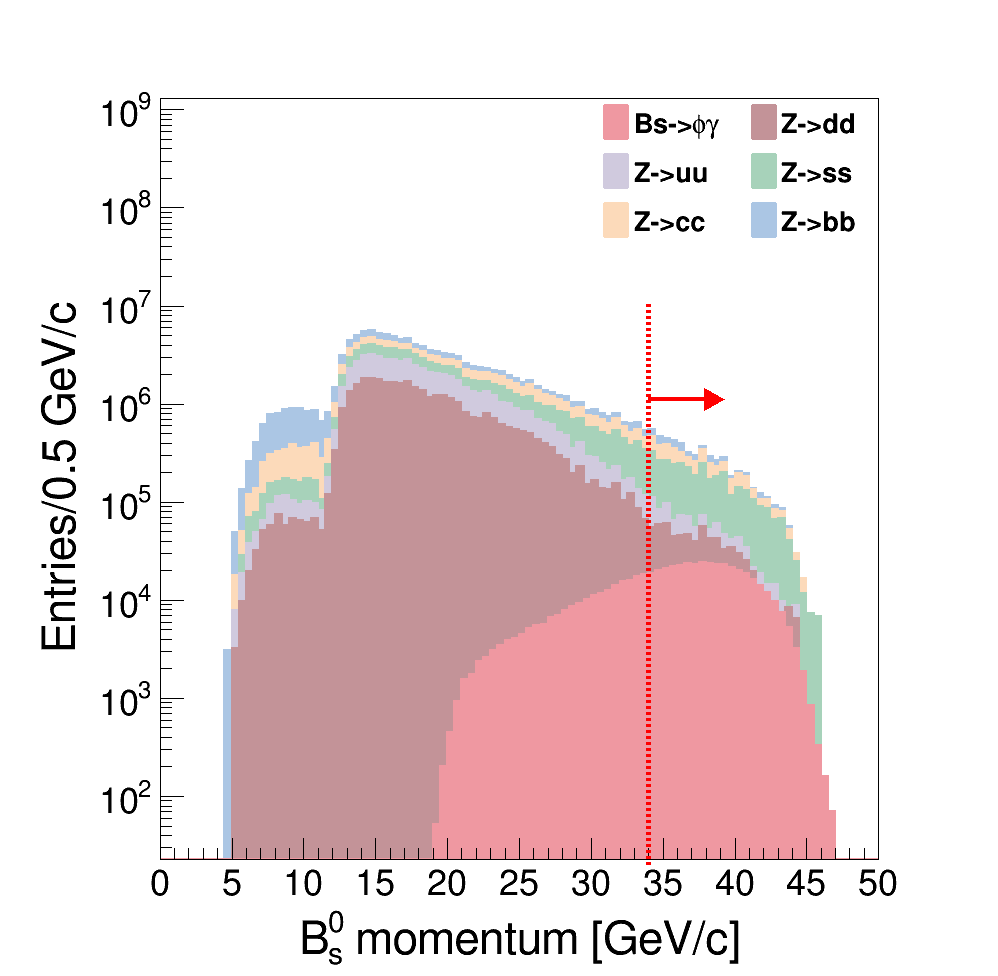}
\caption{Signal and background distributions for $\theta_{\phi\gamma}$ and $P_{B_s^0}$. The selected ranges are $\theta_{\phi\gamma}<0.5$, and $P_{B_s^0}>34\ \mathrm{GeV}/c$.\label{fig:CutChain_Step3}}
\end{figure}

First, we examine the opening angle between the $\phi$ and $\gamma$ momentum, denoted as $\theta_{\phi\gamma}$. Due to the large kinematic boost, the signal is concentrated at small opening angles, as shown in the left panel of Figure~\ref{fig:CutChain_Step3}. The background, however, also has non-negligible contributions at small opening angles. Therefore, \(\theta_{\phi\gamma}\) is not used as a standalone discriminator, but as part of the full selection chain. By minimizing $\sqrt{S+B}/S$, a selection of $\theta_{\phi\gamma} < 0.5$~rad is applied.

Next, we consider the full momentum of the reconstructed $B_s^0$ candidate, $P_{B_s^0}$. As expected from the hard fragmentation of the $b$-quark, the signal $P_{B_s^0}$ distribution peaks at high momentum values, which is distinctly different from the combinatorial background, as illustrated in the right panel of Figure~\ref{fig:CutChain_Step3}. Since \(\theta_{\phi\gamma}\) and \(P_{B_s^0}\) can be correlated, the sequential one-dimensional selections are used here as a transparent cut-based baseline. The multidimensional optimization of these and other variables is further addressed by the BDT analysis in Section~\ref{sec:BDT}. We minimize $\sqrt{S+B}/S$ and apply a selection of $P_{B_s^0} > 34$~$\mathrm{GeV}/c$.

\begin{figure}[h]
    \centering
    \includegraphics[width=.45\textwidth]{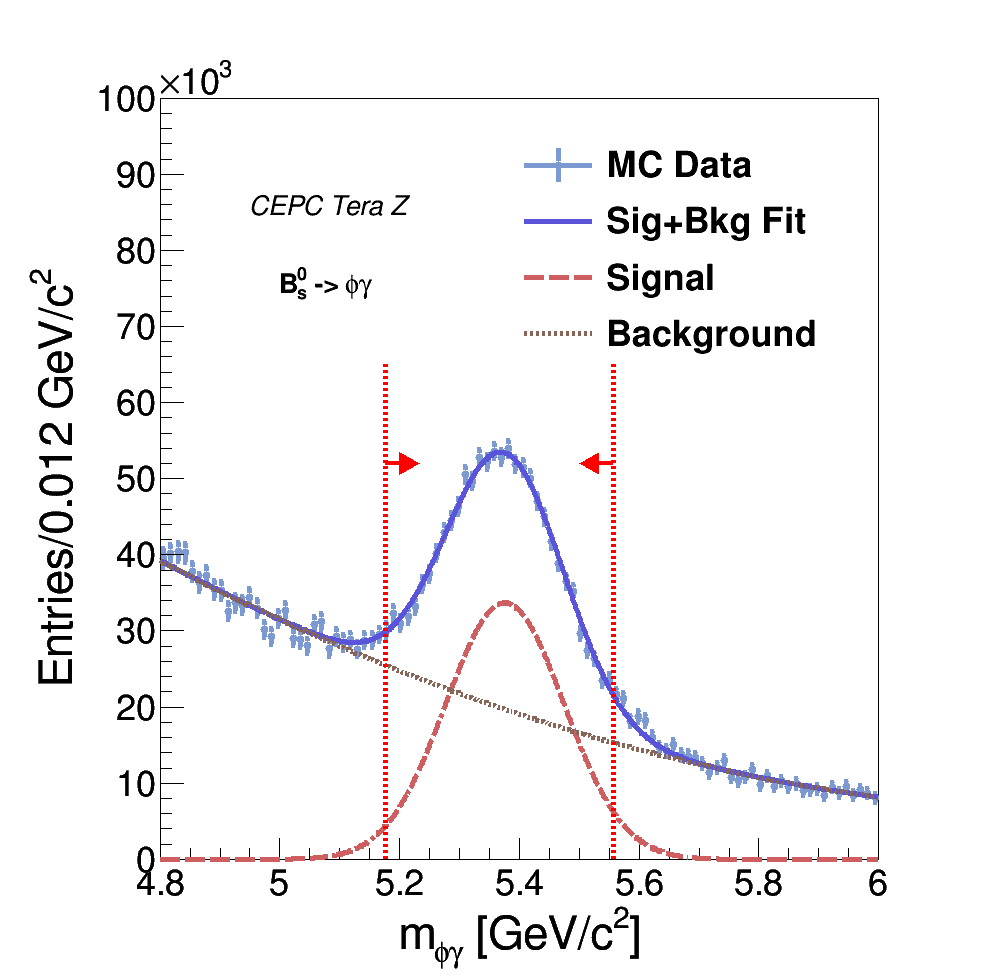}
    \caption{The invariant mass distribution of the reconstructed $B_s^0$ candidates prior to $b$-tagging. The optimal relative statistical uncertainty at this stage is $(0.2210 \pm \text{0.0030})\%$. Note that the data shown are derived from MC samples scaled to $4 \times 10^{12}$ $Z$ bosons (Tera-$Z$); the actual generated MC events correspond to approximately 1/3000 of the displayed counts. \label{fig:Bs0mass_before}}
\end{figure}

Finally, we analyze the invariant mass of the $\phi\gamma$ system, $m_{\phi\gamma}$. The $m_{\phi\gamma}$ invariant mass distribution is shown in Figure~\ref{fig:Bs0mass_before}. To optimize the signal selection, we apply a mass window of $5.177 < m_{\phi\gamma} < 5.557$~$\mathrm{GeV}/c^2$. Within this region, the optimal relative statistical uncertainty reaches $(0.2210 \pm \text{0.0030})\%$. The fitted $B_s^0$ mass peaks at $5.375$~$\mathrm{GeV}/c^2$, with a mass resolution of $99.0$~$\mathrm{MeV}/c^2$. At this stage, the reconstruction efficiency and purity are $(30.50 \pm \text{0.13})\%$ and $(32.1 \pm \text{1.7})\%$, respectively.

\subsubsection{$b$-tagging}

Following the kinematic selections, to further suppress the remaining light-flavor and charm backgrounds, a $b$-tagging selection is applied. As illustrated in Figure~\ref{fig:topology}, in di-jet events, one jet is reconstructed as the signal candidate, while the recoiling jet is used for $b$-tagging. 

\begin{figure}[h]
    \centering
    \includegraphics[width=.42\textwidth]{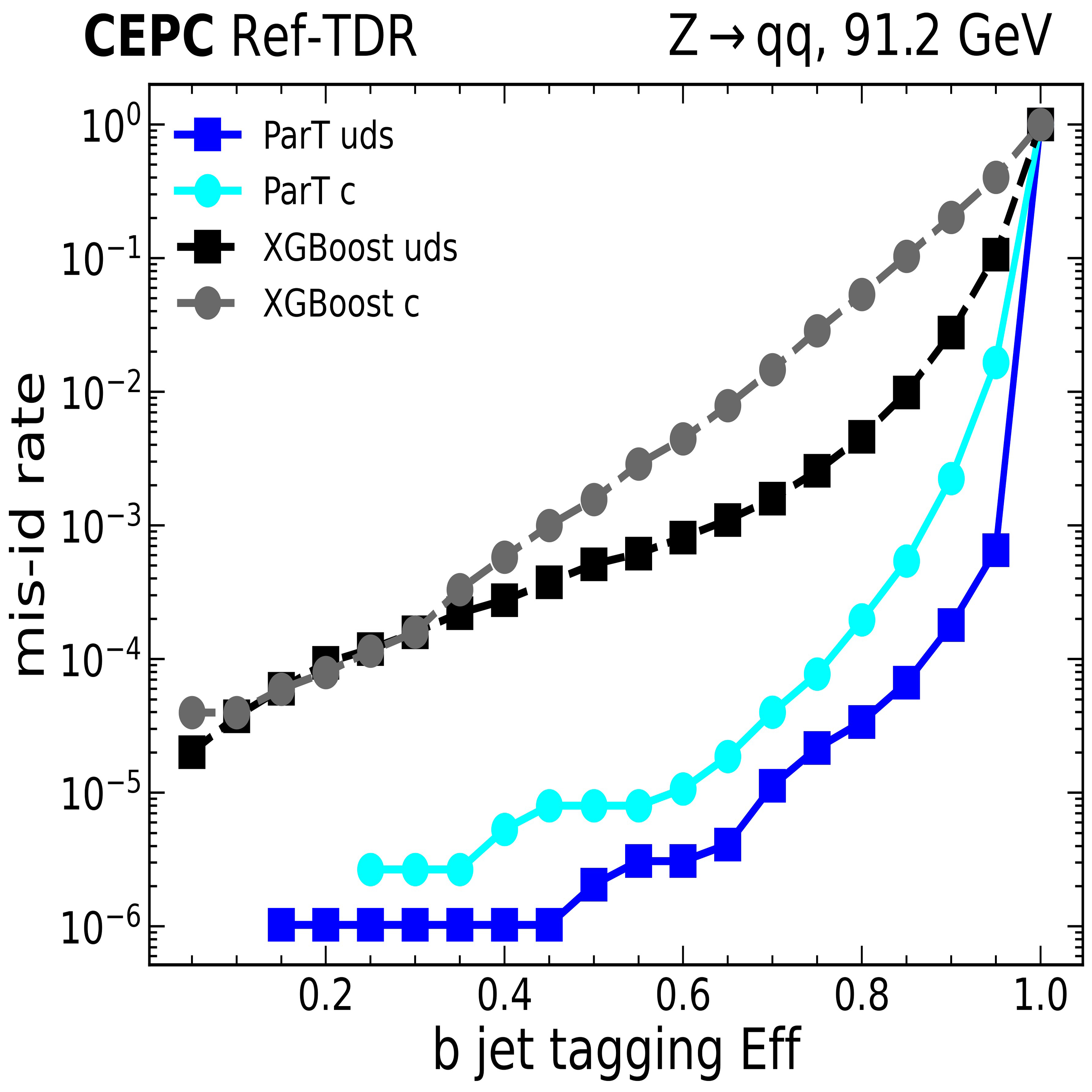}
    \includegraphics[width=.45\textwidth]{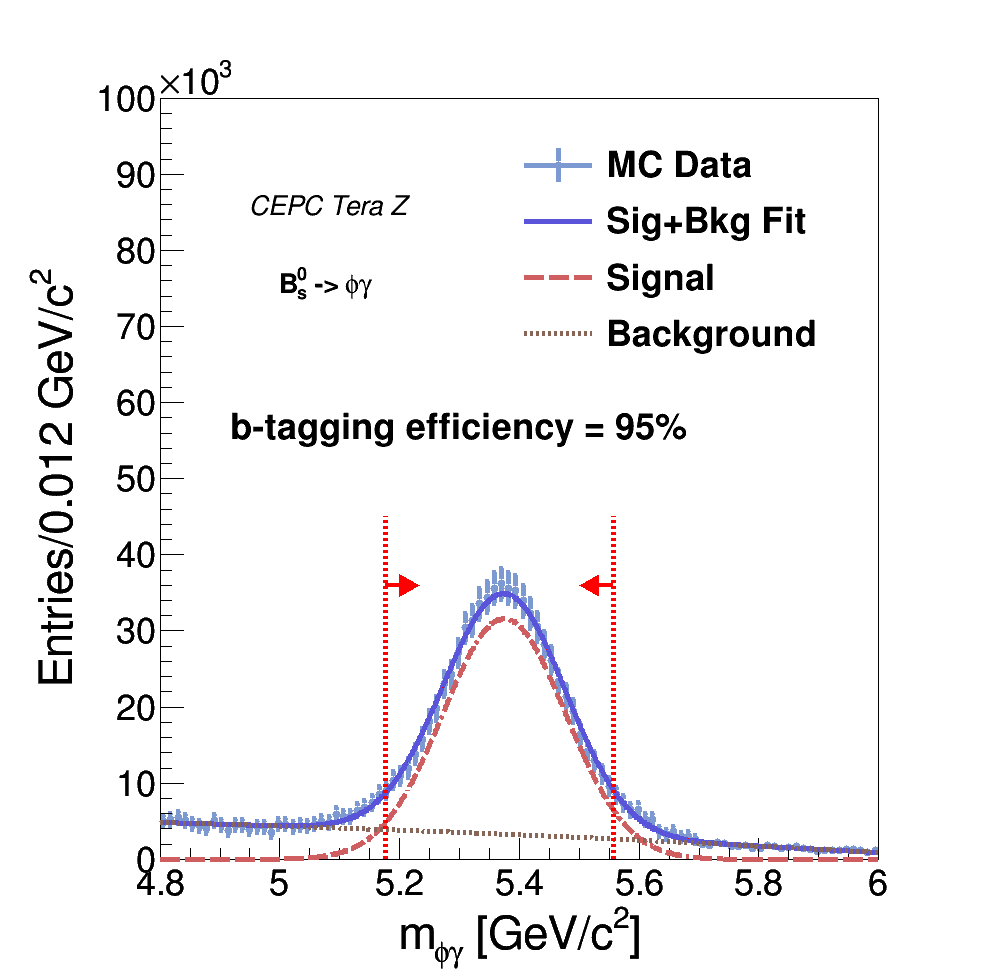}
    \caption{Left: The flavor-tagging performance derived from the CEPC Ref-TDR full simulation studies \cite{Zhang_RefTDR_2025}. Right: The invariant mass distribution of the reconstructed $B_s^0$ candidates after applying the selected $b$-tagging working point ($\epsilon_b = 95\%$). The final relative statistical uncertainty improves to \(0.1384\% \pm 0.0017\%\). The scaling note for the generated MC events is identical to Figure~\ref{fig:Bs0mass_before}. \label{fig:Bs0mass_after}}
\end{figure}

Instead of evaluating a standalone flavor-tagging algorithm directly on our fast-simulation samples, we adopt parametric tagging efficiencies derived from the official CEPC Reference Technical Design Report (Ref-TDR) full-simulation studies \cite{Zhang_RefTDR_2025}, which utilizes the Particle Transformer method \cite{Liang:2023wpt,Qu:2022mxj,Zhu:2025eoe}, as shown in the left panel of Figure~\ref{fig:Bs0mass_after}. Utilizing these official performance metrics ensures a highly realistic estimation of the background rejection capabilities expected for the CEPC detector. We select a high-efficiency working point where the true $b$-jet tagging efficiency is $\epsilon_b = 95\%$, and the corresponding mis-identification rates for charm jets and light-flavor jets are $\epsilon_{c \to b} = 1.66\%$ and $\epsilon_{dus \to b} = 0.065\%$, respectively. By matching the Monte Carlo truth information of the initial quark flavors, these efficiencies are applied as survival probabilities to our remaining simulated events.

\begin{table}[h]
    \centering
    \caption{Event selection results for signal and background candidates %{\color{blue} We don't need log XX since all numbers are small.}{\color{red}The Chi2 value and the reconstructed vertex position have a wide range at X Axis. Additionally, using log XX  enables a more clear of the difference in the distribution of signals and background. So i think log XX is necessary}
    }
    \label{tab:cutchain}
    \adjustbox{max width=\textwidth}{
    \begin{tabular}{>{\centering\arraybackslash}m{4.5cm}>{\centering\arraybackslash}m{1.8cm}>{\centering\arraybackslash}m{1.8cm}>{\centering\arraybackslash}m{1.8cm}>{\centering\arraybackslash}m{1.8cm}>{\centering\arraybackslash}m{2.0cm}>{\centering\arraybackslash}m{2.5cm}}
    \toprule
        Events & Signal & \multicolumn{3}{c}{Background} & $S/(S+B)$ & Rel. Stat. Unc.(\%)\\[8pt]
        \hline
        Event selection & $B_s \to \phi( \to K^+K^-)\gamma$ & $Z \to d\bar{d},u\bar{u},s\bar{s}$ & $Z \to c\bar{c}$ & Remaining $Z \to b\bar{b}$ &  & $\sqrt{S + B}/S$ \\[8pt]
                                                            & $2.08 \times 10^6$ & $1.76 \times 10^{12}$ & $4.90 \times 10^{11}$ & $6.20 \times 10^{11}$ & $7.25 \times 10^{-7}$ & $81.45\pm 0.15$ \\[8pt]
        \hline
        $K^+K^-$ Pair Selection                             & $2.07 \times 10^6$ & $6.37 \times 10^{11}$ & $2.38 \times 10^{11}$ & $3.66 \times 10^{11}$ & $1.67 \times 10^{-6}$ & $53.82\pm0.13$ \\[8pt]
        $\log10(\chi^2)<1$                                  & $2.00 \times 10^6$ & $5.64 \times 10^{11}$ & $1.91 \times 10^{11}$ & $2.43 \times 10^{11}$ & $2.00 \times 10^{-6}$ & $49.95\pm0.11$ \\[8pt]
        $\log10(V_{\phi\to KK}/\mu m)>2.6$                  & $1.55 \times 10^6$ & $6.33 \times 10^{10}$ & $4.81 \times 10^{10}$ & $1.39 \times 10^{11}$ & $6.19 \times 10^{-6}$ & $32.28\pm0.07$ \\[8pt]
        $1.011<m_{K^+K^-}<1.027\ \mathrm{GeV}/c^2$                       & $1.37 \times 10^6$ & $7.39 \times 10^9$    & $8.24 \times 10^9$    & $2.16 \times 10^{10}$ & $3.68 \times 10^{-5}$ & $14.083\pm0.034$ \\[8pt]
        \hline
        $\theta_{\phi\gamma}<0.5\ rad$                           & $1.10 \times 10^6$ & $6.67 \times 10^7$    & $1.62 \times 10^7$    & $1.64 \times 10^7$    & $1.10 \times 10^{-2}$ & $0.911\pm0.006$ \\[8pt]
        $P_{B_s^0}>34\ \mathrm{GeV}/c$                                 & $7.12 \times 10^5$ & $2.13 \times 10^6$    & $8.51 \times 10^5$    & $5.46 \times 10^5$    & $1.68 \times 10^{-1}$ & $0.289\pm0.004$ \\[8pt]
        $5.177<m_{\phi\gamma}<5.557\ \mathrm{GeV}/c^2$                   & $6.38 \times 10^5$ & $1.05 \times 10^6$    & $2.02 \times 10^5$    & $9.92 \times 10^4$    & $3.21 \times 10^{-1}$ & $0.2210\pm0.0030$ \\[8pt]
        \hline
        b-tagging($\epsilon = 95\%, 1.66\%, 0.065\%$)       & $6.06 \times 10^5$ & $2.05 \times 10^2$ & $3.36 \times 10^2$ & $9.42 \times 10^4$ & $8.65 \times 10^{-1}$ & $0.1384\pm0.0017$ \\[8pt]
    \bottomrule
    \end{tabular}
    }
\end{table}
After applying the $b$-tagging survival probabilities, the background is dramatically suppressed by approximately 94\%, while the signal loss is only 5\%. The invariant mass distribution of the reconstructed $B_s^0$ candidates after $b$-tagging is shown in the right panel of Figure~\ref{fig:Bs0mass_after}. The fitted $B_s^0$ mass peaks at $5.374$~$\mathrm{GeV}/c^2$ with a mass resolution of $98.0$~$\mathrm{MeV}/c^2$. Within the selected \(B_s^0\) invariant mass window \(5.177 < m_{\phi\gamma} < 5.557\ \mathrm{GeV}/c^2\), the optimal measurement relative statistical uncertainty reaches \(0.1384\% \pm 0.0017\%\), with corresponding efficiency and purity of $(29.13 \pm \text{0.13})\%$ and $(86.1 \pm \text{2.0})\%$. The complete selection process is summarized in Table~\ref{tab:cutchain}.

\subsection{Multivariate analysis with BDT}\label{sec:BDT}

\begin{table}[h]
    \centering
    \caption{The event pre-selection results of signal and background for BDT }
    \label{tab:cutchain_BDT}
    \adjustbox{max width=\textwidth}{
    \begin{tabular}{>{\centering\arraybackslash}m{4.5cm}>{\centering\arraybackslash}m{1.8cm}>{\centering\arraybackslash}m{1.8cm}>{\centering\arraybackslash}m{1.8cm}>{\centering\arraybackslash}m{1.8cm}>{\centering\arraybackslash}m{2.0cm}>{\centering\arraybackslash}m{2.5cm}}
    \toprule
        Events & Signal & \multicolumn{3}{c}{Background} & $S/(S+B)$ & Rel. Stat. Unc.(\%)\\[8pt]
        \hline
        Event selection & $B_s \to \phi( \to K^+K^-)\gamma$ & $Z \to d\bar{d},u\bar{u},s\bar{s}$ & $Z \to c\bar{c}$ & Remaining $Z \to b\bar{b}$ &  & $\sqrt{S + B}/S$ \\[8pt]
                                                            & $2.08 \times 10^6$ & $1.76 \times 10^{12}$ & $4.90 \times 10^{11}$ & $6.20 \times 10^{11}$ & $7.25 \times 10^{-7}$ & $81.45\pm0.15$\\[8pt]
        \hline
        $K^+K^-$ Pair Selection                             & $2.07 \times 10^6$ & $6.37 \times 10^{11}$ & $2.38 \times 10^{11}$ & $3.66 \times 10^{11}$ & $1.67 \times 10^{-6}$ & $53.82\pm0.13$\\[8pt]
        $\log10(\chi^2)<1$                                  & $2.00 \times 10^6$ & $5.64 \times 10^{11}$ & $1.91 \times 10^{11}$ & $2.43 \times 10^{11}$ & $2.00 \times 10^{-6}$ & $49.95\pm0.11$\\[8pt]
        $\log10(V_{\phi\to KK}/\mu m)>2.6$                  & $1.55 \times 10^6$ & $6.33 \times 10^{10}$ & $4.81 \times 10^{10}$ & $1.39 \times 10^{11}$ & $6.19 \times 10^{-6}$ & $32.28\pm0.07$\\[8pt]
        $1.011<m_{K^+K^-}<1.027\ \mathrm{GeV}/c^2$                       & $1.37 \times 10^6$ & $7.39 \times 10^9$    & $8.24 \times 10^9$    & $2.16 \times 10^{10}$ & $3.68 \times 10^{-5}$ & $14.08\pm0.034$\\[8pt]
        $\theta_{\phi\gamma}<1.5\ rad$                           & $1.48 \times 10^6$ & $1.40 \times 10^{9}$  & $1.20 \times 10^{9}$  & $1.38 \times 10^{9}$  & $3.72 \times 10^{-4}$ & $4.26\pm0.011$ \\[8pt]
        $P_{B_s^0}>10\ \mathrm{GeV}/c$                                 & $1.47 \times 10^6$ & $1.24 \times 10^9$    & $9.97 \times 10^8$    & $7.90 \times 10^8$    & $4.85 \times 10^{-4}$ & $3.741\pm0.010$  \\[8pt]
        $5.0<m_{\phi\gamma}<5.7\ \mathrm{GeV}/c^2$                       & $1.41 \times 10^6$ & $2.10 \times 10^8$    & $3.89 \times 10^8$    & $2.98 \times 10^8$    & $1.57 \times 10^{-3}$ & $2.1362\pm0.0065$ \\[8pt]
        \hline
        b-tagging($\epsilon = 95\%, 1.66\%, 0.065\%$)       & $1.34 \times 10^6$ & $3.24 \times 10^5$    & $6.46 \times 10^6$    & $2.82 \times 10^8$    & $4.62 \times 10^{-3}$ & $1.270\pm0.005$ \\[8pt]
        \hline
        BDT responses larger than 0.18                               & $9.43 \times 10^5$ & $299$ & $5572$ & $6.19 \times 10^4$ & $9.33 \times 10^{-1}$ & $0.107\pm0.004$\\[8pt]
    \bottomrule
    \end{tabular}
    }
\end{table}

\begin{figure}[h]
    \centering
    \includegraphics[width=0.55\textwidth]{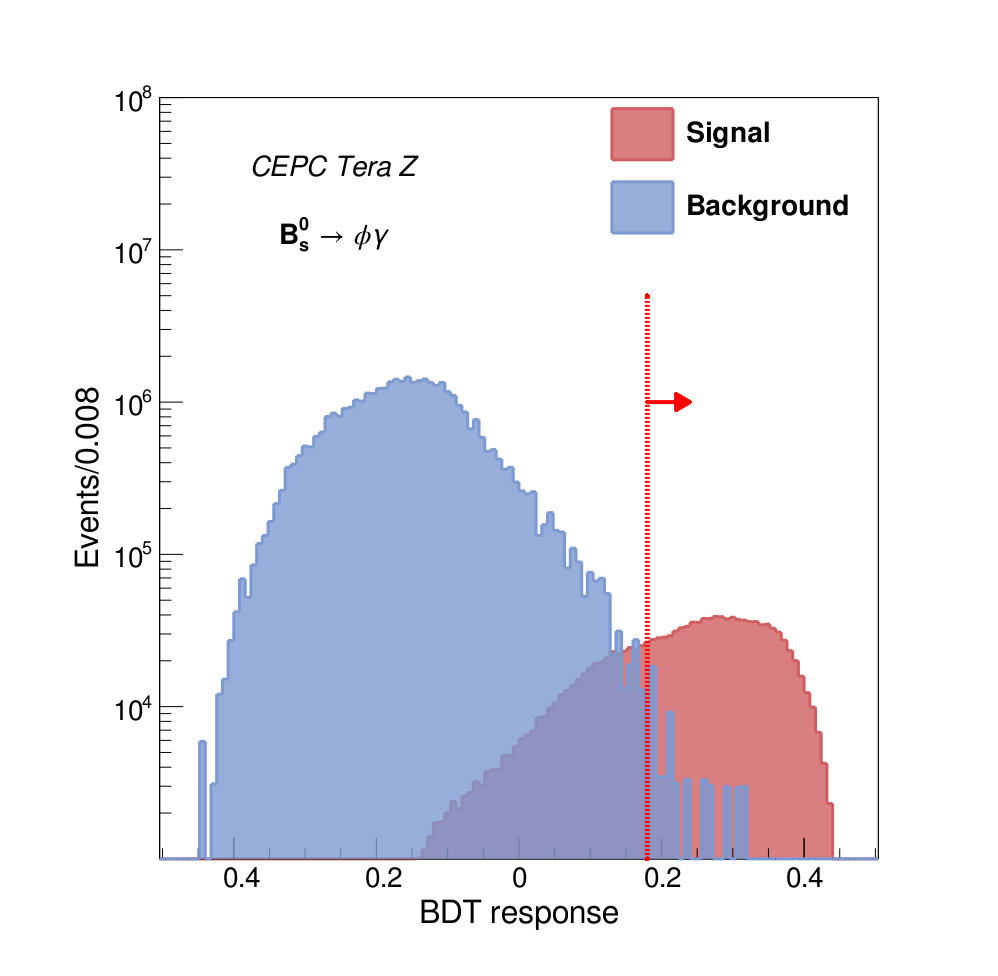}
    \caption{The BDT response distributions for the signal and the background after the pre-selection. The detailed pre-selection criteria are listed in Table~\ref{tab:cutchain_BDT}. The vertical dashed line indicates the optimized BDT responses larger than 0.18.\label{fig:BDT}}
\end{figure}

While the cut-based optimization demonstrates excellent background suppression, it relies on sequential rectangular selections. Such an approach may be suboptimal when dealing with multi-dimensional phase spaces where kinematic variables exhibit strong non-linear correlations. To fully exploit these multi-dimensional features and further enhance the signal-to-background separation, a multivariate analysis utilizing a BDT algorithm is employed.

To provide the BDT with a sufficiently large training sample while effectively rejecting backgrounds, a loosened pre-selection chain is applied before the BDT training. Specifically, the pre-selection criteria are defined as follows: $\log_{10}(\chi^2)<1$, $\log_{10}(V_{\phi\to K^+K^-}/\mu\text{m})>2.6$, $1.01<m_{K^+K^-}<1.07$~$\mathrm{GeV}/c^2$, $\theta_{\phi\gamma}<1.5$~rad, $P_{B_s^0}>10$~$\mathrm{GeV}/c$, and a widened $B_s^0$ mass window of $5.0<m_{\phi\gamma}<5.7$~$\mathrm{GeV}/c^2$. Following this, the standard $b$-tagging survival probabilities are applied. The detailed event yields after this pre-selection are summarized in Table~\ref{tab:cutchain_BDT}.

The pre-selected events then serve as inputs for the BDT training. The six kinematic variables mentioned in the pre-selection are used as the training features. Figure~\ref{fig:BDT} illustrates the BDT response distributions for both the signal and the remaining combinatorial background. By minimizing $\sqrt{S+B}/S$, events with BDT responses larger than 0.18 are retained. The remaining signal and background event yields are $9.4 \times 10^{5}$ and $6.8 \times 10^{4}$, respectively. The final achieved selection efficiency and purity are $(45.3 \pm \text{0.2})\%$ and $(93.3 \pm \text{1.6})\%$. Consequently, the expected relative statistical uncertainty reaches $(0.107 \pm 0.004)\%$, representing an approximately 22\% improvement over the cut-based method and an enhancement of roughly two orders of magnitude compared to current experimental results \cite{LHCb:2012quo}.

\section{Impact of mixing-induced and CP-violating observables}
% \section{Impact of $Bs/\bar{Bs}$ mixed oscillation}

\subsection{Impact of $B_s^0$-$\bar{B}_s^0$ oscillation}
In the SM, photons in the \(b \to s\gamma\) process are expected to be purely left-handed, a consequence of the W boson’s coupling to left-handed quarks (or right-handed antiquarks). However, NP contributions could introduce right-handed helicity photons. While challenging to measure directly, these contributions can be probed indirectly through time-dependent studies of $B_s^0$ and $\bar{B_s^0}$ radiative decays \cite{PhysRevLett.79.185}.

In \(B_s^0 \to \phi\gamma\) decays, the interference between \(B_s^0\) mixing and decay amplitudes modulates the decay rate as a function of proper time \cite{Peng:2025bki}. This time dependence is described by \cite{MUHEIM2008174}:
\begin{align}
&\mathcal{P}(t) \propto e^{-\Gamma_s t}\left\{ \cosh\left(\frac{\Delta\Gamma_s t}{2}\right) - \mathcal{A}^\Delta \sinh\left(\frac{\Delta\Gamma_s t}{2}\right) + \zeta \mathcal{C} \cos(\Delta m_s t) - \zeta \mathcal{S} \sin(\Delta m_s t) \right\} \label{eq:1}
\end{align} where \(\Delta\Gamma_s = 0.082\pm 0.005\ \text{ps}^{-1}\) \cite{ParticleDataGroup:2024cfk} and \(\Delta m_s = 17.765\pm 0.005\ \text{ps}^{-1}\) \cite{ParticleDataGroup:2024cfk} are the width and mass differences between the \(B_s^0\) mass eigenstates, \(\Gamma_s = 0.6598\pm 0.0025\ \text{ps}^{-1}\) \cite{ParticleDataGroup:2024cfk} is the mean decay width, $\zeta = +1$ $(-1)$ corresponds to an initial $B_s^0$ ($\bar{B}_s^0$). The coefficients \(\mathcal{A}^\Delta\), \(\mathcal{C}\) and \(\mathcal{S}\) are functions of the left and right handed photon polarization amplitudes \cite{MUHEIM2008174}. The coefficients \(\mathcal{A}^\Delta\) and \(\mathcal{S}\) are sensitive to the photon helicity amplitudes and weak phases, while \(\mathcal{C}\) characterizes direct CP violation. The SM predicts small values for these parameters \(\mathcal{A}_{\text{SM}}^\Delta = 0.047_{-0.025}^{+0.029}\), \(\mathcal{S}_{\text{SM}} = 0 \pm 0.002\), and \(\mathcal{C}_{\text{SM}} \approx 0.005 \pm 0.005\) \cite{MUHEIM2008174}, reflecting the expected minimal CP violation and dominant left-handed photon helicity in the SM. The LHCb collaboration has measured these parameters using approximately 5000 \(B_s^0 \to \phi\gamma\) signal events, yielding \cite{PhysRevLett.123.081802,PhysRevLett.118.021801}: \(\mathcal{S}_{\phi\gamma} = 0.43 \pm 0.30 \pm 0.11\), \(\mathcal{C}_{\phi\gamma} = 0.11 \pm 0.29 \pm 0.11\), and \(\mathcal{A}_{\phi\gamma}^\Delta = -0.67_{-0.41}^{+0.37} \pm 0.17\), where the first uncertainty is statistical and the second systematic. However, LHCb did not observe a clear oscillatory pattern in the decay-time distribution, which limits the precision of the extracted CP-violating observables. In contrast, the clean \(e^+e^-\) environment at the CEPC, together with efficient background suppression and well-controlled detector conditions, is expected to improve the measurement precision of \(A^\Delta\), \(C\), and \(S\). 

\begin{figure}[h]
\centering
\includegraphics[width=0.5\textwidth]{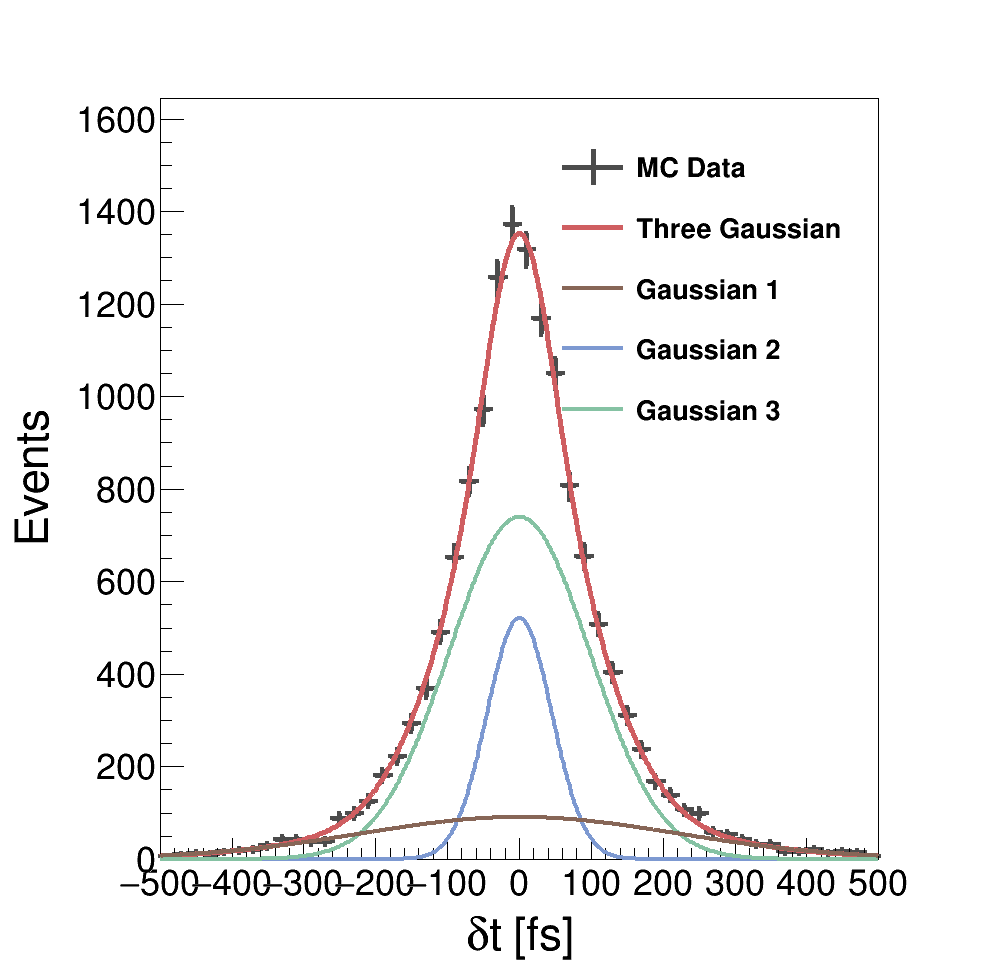}
\caption{The reconstructed $B_s^0$ lifetime resolution distribution from full simulation, modeled as a mixture of three Gaussian functions achieving an effective resolution of $\sigma_t = 76fs$. %{\color{blue}{How's the $B_s$ momentum reconstructed to obtain lifetime? Is the photon energy (and $K$ energy) corrected by any methods?}}{\color{red}{1. The lifetime of $B_s^0$ mesons is derived from the reconstructed vertex position and transverse momentum of the $B_s^0$ candidate, following the relation $\tau = \frac{m_sl_{xy}}{p_T}$}. 2. The energy of gamma and $KK$ is obtained through the full simulation process, by directly reading the information from ArborPFO. {\color{blue} I simply wonder if there is anyway to improve the resolution from with kinematic informations. Sounds like a topic for a future study. Also, why can't we use the physical mass of $B_s$ instead of $m_s$? It has almost no uncertainty.}{\color{red}1. Indeed, if the time resolution can be improved, then we will observe a more significant $B_s$ oscillation (the smear from the green line to the gray line will be smaller in Figure 11). 2. I am sorry I have made a mistake, the $m_s$ used here refers to the "truth Mass" of $B_s$, rather than the reconstructed mass.}}
\label{fig:TimeReso}}
\end{figure}

To precisely measure these time-dependent CP-violating observables, an initial-state flavor-tagging algorithm \cite{Peng:2025bki} is applied to distinguish whether the signal meson was initially produced as a $B_s^0$ or a $\bar{B}_s^0$. It is important to distinguish this procedure from the $b$-jet tagging for background suppression used in Section 3.1.3, which merely identifies the generic presence of a $b$-quark rather than its specific charge sign. The algorithm categorizes the reconstructed $B_s^0$ candidates into correctly tagged ($R$), incorrectly tagged ($W$), and untagged ($U$) events. The initial-state flavor-tagging efficiency is $\epsilon_{\text{tag}} = \frac{R+W}{R+W+U} = 72.1\%$. The mis-tag fraction is calculated as $\omega = \frac{W}{R+W} = 21.4 \%$. The effective tagging efficiency defined as $\epsilon_{eff} = \epsilon_{tag}(1-2\omega)^2$ , is calculated to be 23.6\% \cite{Peng:2025bki}.

The lifetime of \(B_s^0\) mesons is derived from the reconstructed vertex position and transverse momentum of the \(B_s^0\) candidate, following the relation:
\begin{align}
&t = \frac{m_sl_{xy}}{p_T} \label{eq:2}
\end{align} 
where $t$ is the reconstructed proper decay time, $m_s$ is the invariant mass of $B_s^0$, $p_T$ is the transverse momentum, and $l_{xy}$ is the projected decay length in the transverse plane, measured specifically as the distance between the interaction point and the decay vertex of the $B_s^0$ meson. 

The distribution of the reconstructed proper time residual, defined as $\delta t = t_{\text{reco}} - t_{\text{truth}}$, is shown in Figure~\ref{fig:TimeReso}, where $t_{\text{reco}}$ and $t_{\text{truth}}$ denote the reconstructed and the Monte Carlo truth decay times, respectively. The reconstructed time is obtained from the full simulation. The nominal value of the decay time resolution is derived from a fit using three Gaussian functions, which are combined into an effective resolution of $\sigma_t = 76$~fs following the function:
\begin{align}
&\sigma_t = \sqrt{-\frac{2}{\Delta m_s^2}\ln{f_1e^{-\frac{1}{2}\sigma_1^2\Delta m_s^2} + f_2e^{-\frac{1}{2}\sigma_2^2\Delta m_s^2} + f_3e^{-\frac{1}{2}\sigma_3^2\Delta m_s^2}}} \label{eq:3}
\end{align} where  \(f_{1,2,3}\) are the fractional contributions of each Gaussian component, and \(\sigma_{1,2,3}\) are their respective widths. The reconstruction efficiency depends on the $B_s^0$ decay time. Events with decay times that are either too short or too long exhibit reduced efficiencies. This time-dependent efficiency is parameterized by an acceptance function:
\begin{align}
&A(t) = \frac{\alpha t^\beta}{1+\alpha t^\beta} \times(1-\xi t)  \label{eq:4}
\end{align} where the parameters $\alpha = 4.6$, $\beta = 0.89$, and $\xi = 0.088$, are obtained by fitting selected signal MC events.

\begin{figure}[h]
\centering
\includegraphics[width=0.5\textwidth]{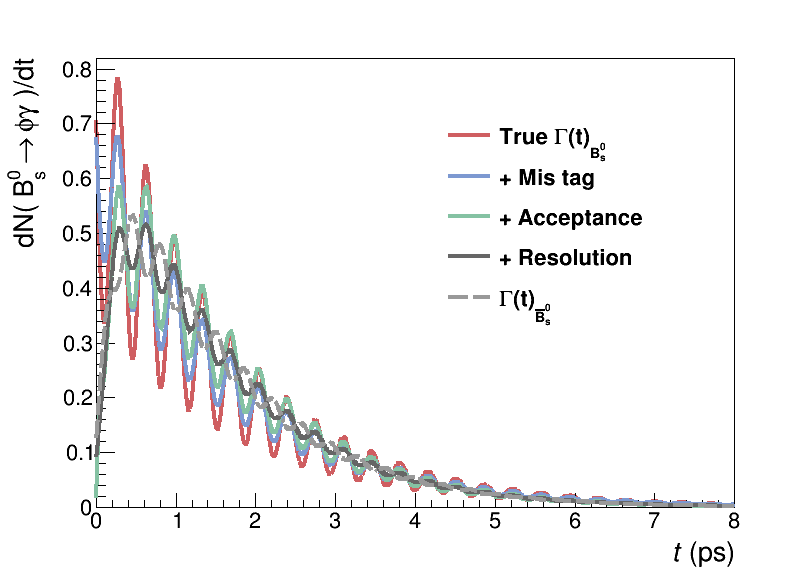}
\caption{The decay time distribution of \(B_{s}^{0} \to \phi \gamma\) at CEPC. %{\color{blue} The y-axis labeling shall be A.U. or $d N(B_s^0\to \phi\gamma) /dt$.}{\color{red} This figure was updated.}
}\label{fig:TimeFunction}
\end{figure}

The expected decay time distribution for $B_{s}^{0} \to \phi\gamma$ at CEPC is illustrated in Figure~\ref{fig:TimeFunction}. This distribution is modeled by a weighted theoretical time-dependent decay rate probability density function (PDF) from Eq.~\ref{eq:1}, convolved with the decay time resolution function and an acceptance function. The red line corresponds to the theoretical distribution of decay time for $B_{s}^{0} $-$\bar{B}_{s}^{0}$ oscillation in the absence of detector effects. After introducing the initial-state flavor-tagging efficiency and the mis-tag fraction into the PDF, the $B_s^0$ decay time distribution corresponds to the blue line in Figure~\ref{fig:TimeFunction}. Subsequently, by convolving with the acceptance function described in Eq.~\ref{eq:4}, we obtain the decay time distribution described by the green line. By further convolving with a Gaussian distribution to describe the time resolution with $\sigma_t = 76$~fs, we obtain the final distribution shown as the black line in Figure~\ref{fig:TimeFunction}. The decay time spectrum of $\bar{B}_s^0$ is depicted by the brown dashed line in Figure~\ref{fig:TimeFunction}. This final distribution represents the expected observed decay time spectrum at the CEPC.

\subsection{Measurement of CP-violating parameters}

The CP-violating parameter $A_{\phi\gamma}^{\Delta}$ can be extracted from untagged $B_s^{0}$ and $\bar{B}_s^{0}$ decays. In untagged samples, the initial flavor of the $B_s^0$ meson is unknown. Consequently, the parameters $C_{\phi\gamma}$ and $S_{\phi\gamma}$ cancel out when averaging over $B_s^0$ and $\bar{B}_s^0$ events. This cancellation eliminates the need for initial-state flavor tagging and simplifies the decay time PDF to a form sensitive only to $A_{\phi\gamma}^{\Delta}$: 
\begin{align}
\mathcal{P}_{\text{untagged}}(t) \propto e^{-\Gamma_s t} \left\{ \cosh\left(\frac{\Delta\Gamma_s t}{2}\right) - A_{\phi\gamma}^{\Delta} \sinh\left(\frac{\Delta\Gamma_s t}{2}\right) \right\} \label{eq:5}
\end{align} 

To validate the extraction procedure for $A_{\phi\gamma}^{\Delta}$ and quantify potential biases and uncertainties, we generate 943K Monte Carlo signal events and 68K background events. These yields correspond exactly to the final selected events obtained from the optimization procedure described in Section 3. The signal events are generated from a combined time-dependent distribution described by Eq.~\ref{eq:5}, with the acceptance function (Eq.~\ref{eq:4}) and time resolution incorporated. The input value is set to $A_{\phi\gamma}^{\Delta} = -0.67$, matching the central value of the LHCb measurement. The background events are sampled from an empirical decay time distribution, described by the function:
\begin{align}
\mathcal{P}_{\text{Bkg}}(t) = 0.516 \exp\left(-\frac{1}{2} \times \left( \frac{\ln(\frac{t}{0.155})}{0.578} \right)^2 \right) + 0.103 \exp\left(-\frac{1}{2} \times \left( \frac{\ln(\frac{t}{0.427})}{0.833} \right)^2 \right) \label{eq:6}
\end{align} 
This empirical function is derived by fitting the decay time distribution of the selected background events described in Section 3. 

\begin{figure}[h]
\centering
\includegraphics[width=0.5\textwidth]{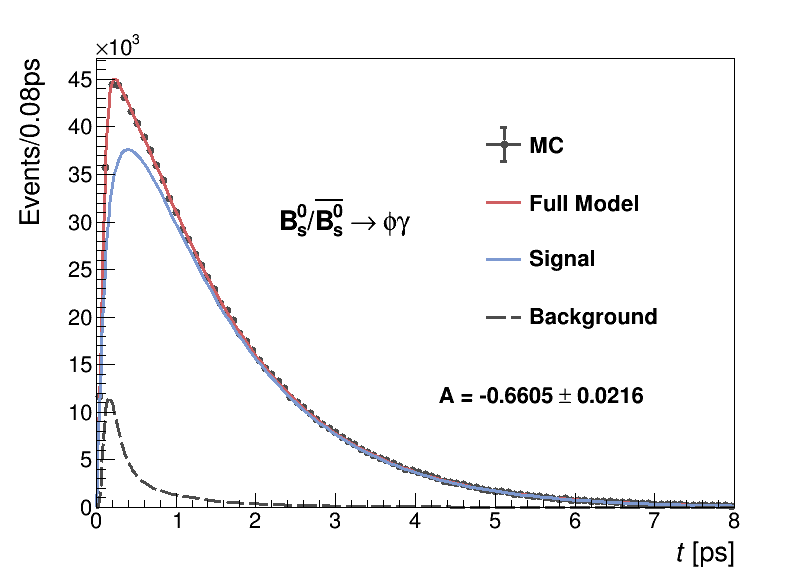}
\caption{The decay time distribution for untagged $B_{s}^{0} \to \phi \gamma$ decays and a single pseudo-experiment fit result for the parameter $A_{\phi\gamma}^{\Delta}$, yielding a fitted value of $A_{\phi\gamma}^{\Delta} = -0.661 \pm 0.022$ with the generated input values of $A_{\phi\gamma}^{\Delta} = -0.67$.\label{fig:FitA}}
\end{figure}

\begin{figure}[h]
\centering
\includegraphics[width=0.4\textwidth]{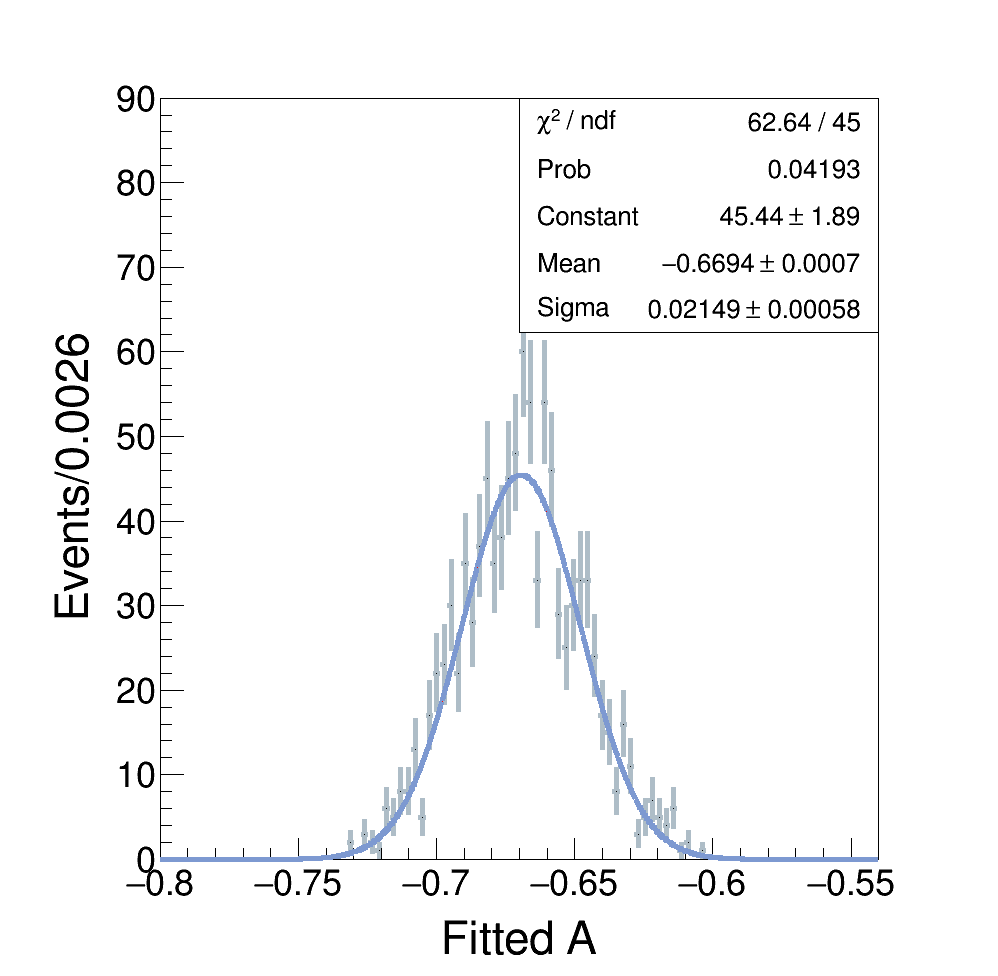}
\includegraphics[width=0.4\textwidth]{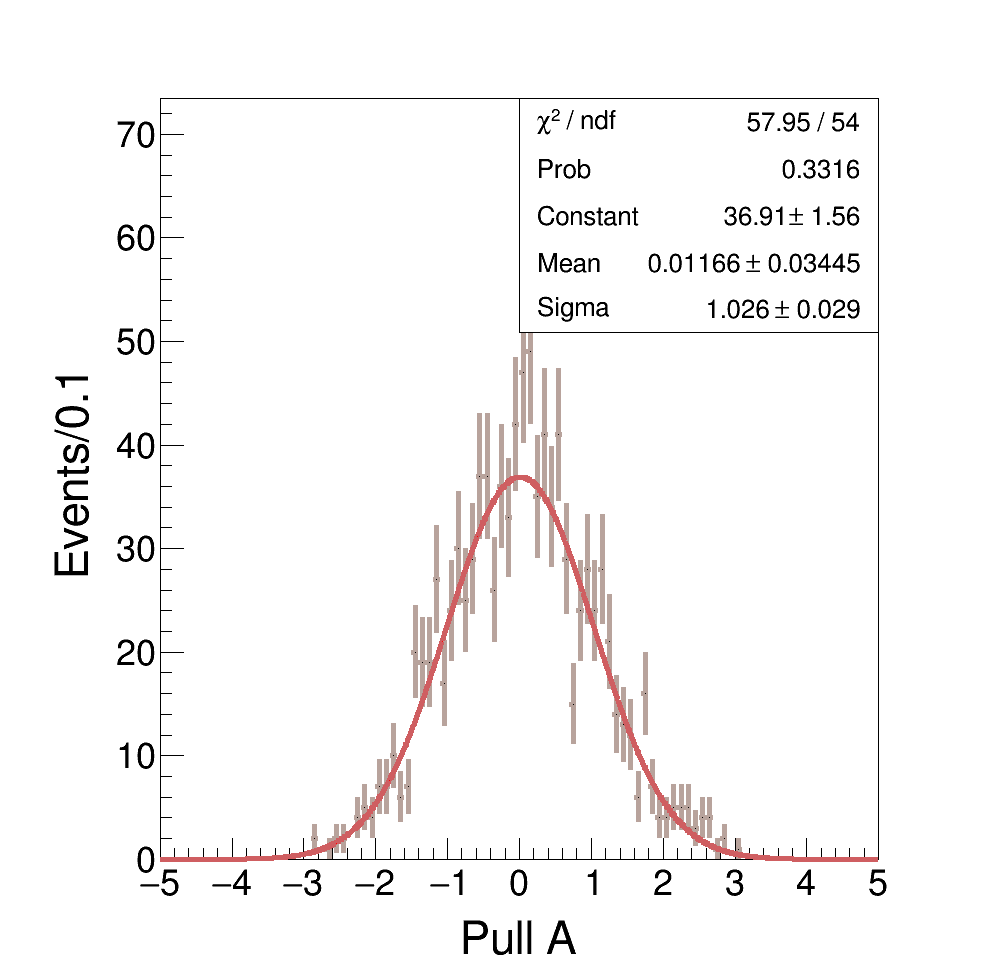}
\caption{The distribution of the fitted value (left) and the pull (right) for the parameter $A_{\phi\gamma}^{\Delta}$, obtained from 1000 pseudo-experiments with an input generated value of $-0.67$. The pull is defined as $\frac{A_{\text{fitted}} - A_{\text{generated}}}{\sigma_A}$.\label{fig:PullA}}
\end{figure}

The decay time distribution of the combined MC signal and background sample, along with the fits, is presented in Figure~\ref{fig:FitA}. The fit accurately models the complex decay time spectrum and successfully extracts the parameters, yielding $A_{\phi\gamma}^{\Delta} = -0.661 \pm 0.022$. We perform 1000 independent pseudo-experiments by repeatedly fitting the combined MC samples. The distribution of the fitted $A_{\phi\gamma}^{\Delta}$ is shown in the left panel of Figure~\ref{fig:PullA}. From this distribution, the mean fitted value is $-0.669$, and the statistical uncertainty per fit is estimated to be $\sigma_{A_{\phi\gamma}^{\Delta}} = 0.021$. The pull quantifies the bias between the fitted result and the true value, and is defined as $\frac{x_{\rm fitted} - x_{\rm generated}}{\sigma_{\rm fitted}}$, where $x_{\rm fitted}$ is the fitted value, $x_{\rm generated}$ is its generated value, and $\sigma_{\rm fitted}$ is the fitting uncertainty of $x_{\rm fitted}$. The distribution of the pull is shown in the right panel of Figure~\ref{fig:PullA}. The central value is 0.012, indicating that the bias of the fitted $A_{\phi\gamma}^{\Delta}$ is well within a reasonable range.

\begin{figure}[h]
\centering
\includegraphics[width=0.5\textwidth]{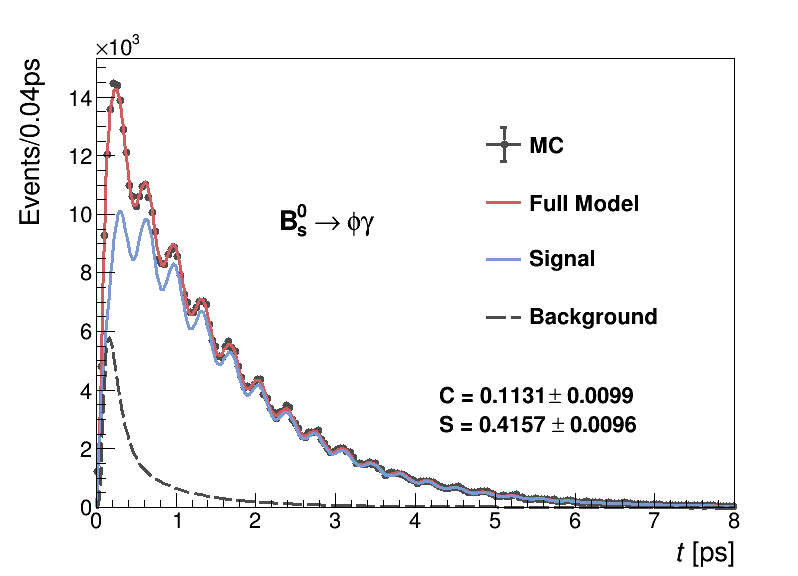}
\caption{The decay time distribution and corresponding fits for tagged $B_{s}^{0} \to \phi \gamma$ events at the CEPC. The extracted parameters are $\mathcal{C}_{\phi\gamma} = 0.113 \pm 0.010$ and $\mathcal{S}_{\phi\gamma} = 0.416 \pm 0.010$ with the generated input values of $\mathcal{C}_{\phi\gamma} = 0.11$ and $\mathcal{S}_{\phi\gamma} = 0.43$, respectively.\label{fig:FitCS}}
\end{figure}
 
\begin{figure}[h]
\centering
\includegraphics[width=0.4\textwidth]{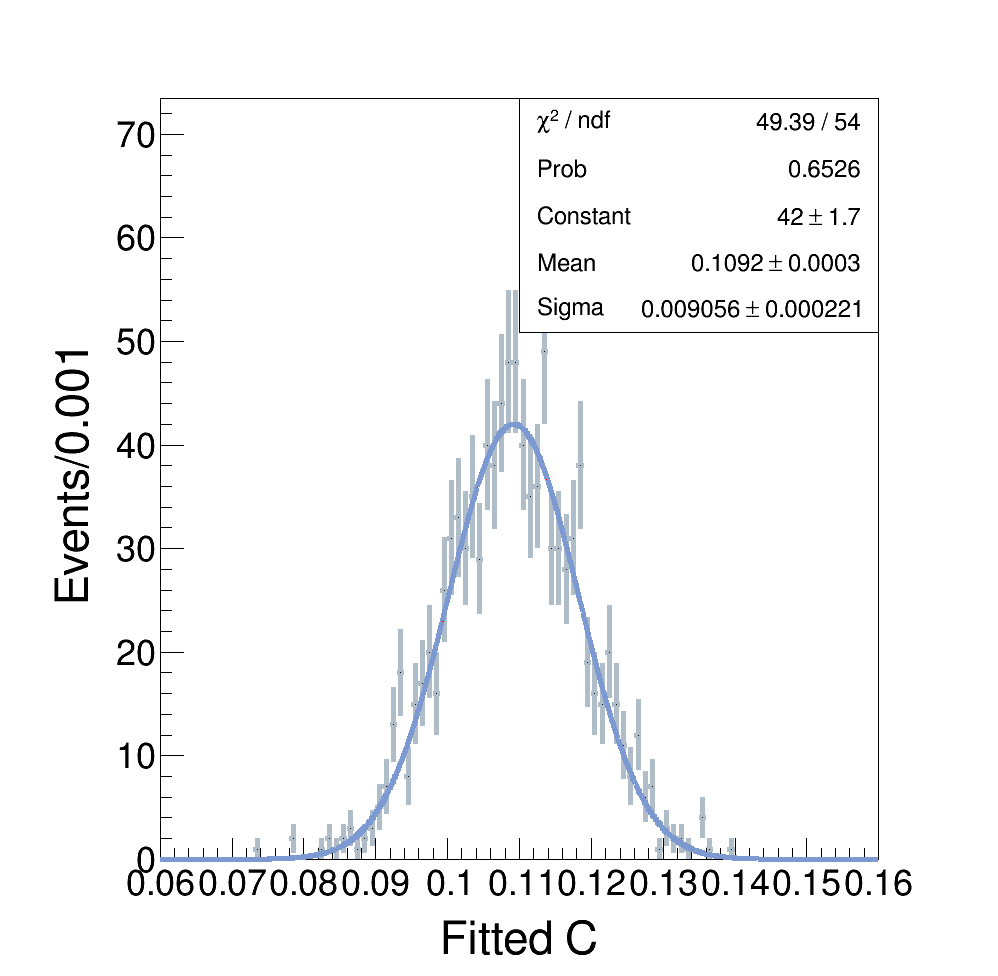}
\includegraphics[width=0.4\textwidth]{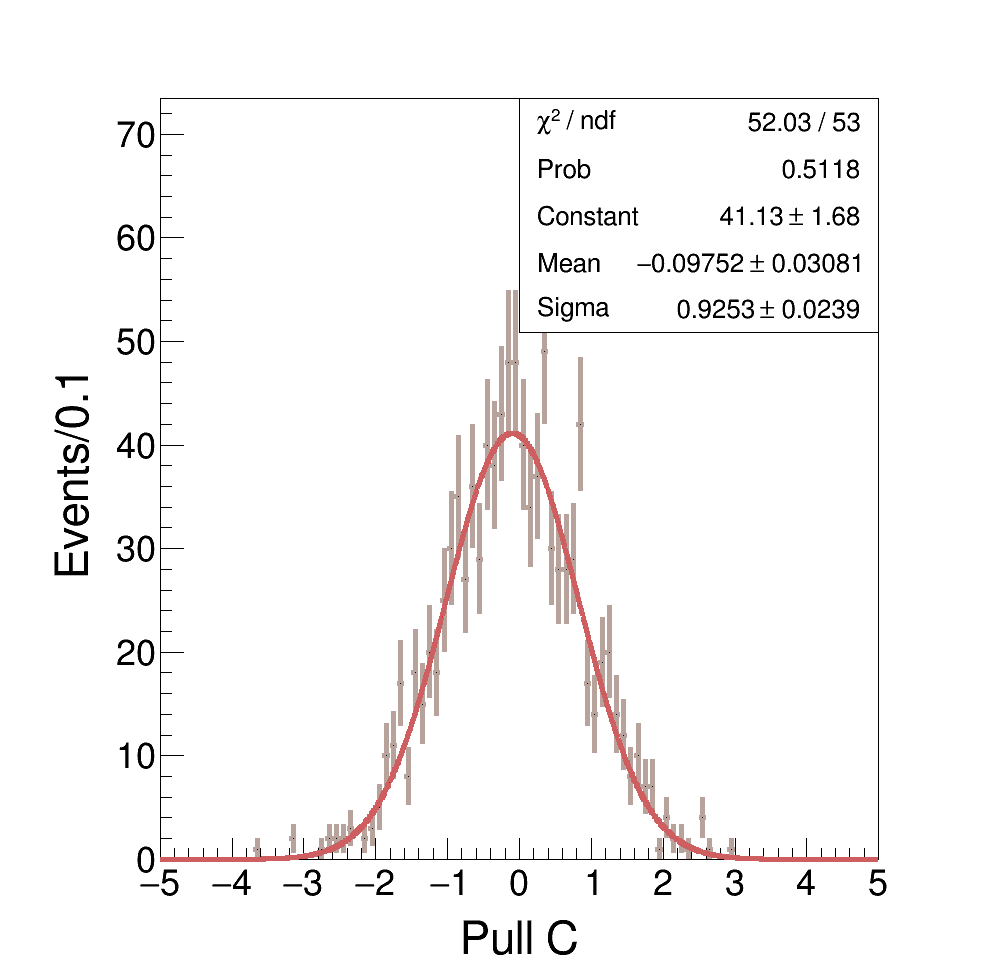}
\caption{Distributions of the fitted value (left) and the pull (right) for \(C_{\phi\gamma}\), obtained from 1000 pseudo-experiments with an input generated value of \(C_{\phi\gamma}^{\mathrm{gen}}=0.11\). The pull is defined as \((C_{\phi\gamma}^{\mathrm{fit}}-C_{\phi\gamma}^{\mathrm{gen}})/\sigma_{C}\), where \(C_{\phi\gamma}^{\mathrm{fit}}\) is the fitted value in each pseudo-experiment, \(C_{\phi\gamma}^{\mathrm{gen}}\) is the generated input value, and \(\sigma_C\) is the fit uncertainty of \(C_{\phi\gamma}^{\mathrm{fit}}\).\label{fig:PullC}}
\end{figure}

\begin{figure}[h]
\centering
\includegraphics[width=0.4\textwidth]{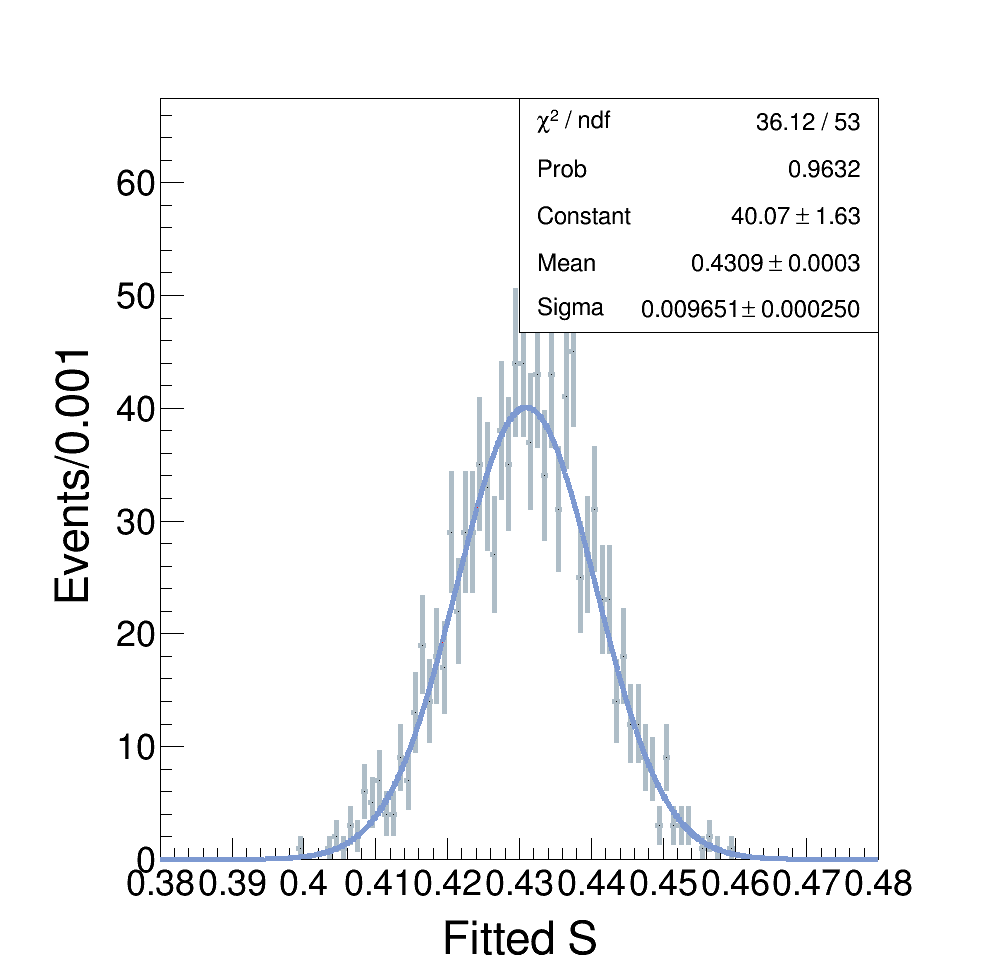}
\includegraphics[width=0.4\textwidth]{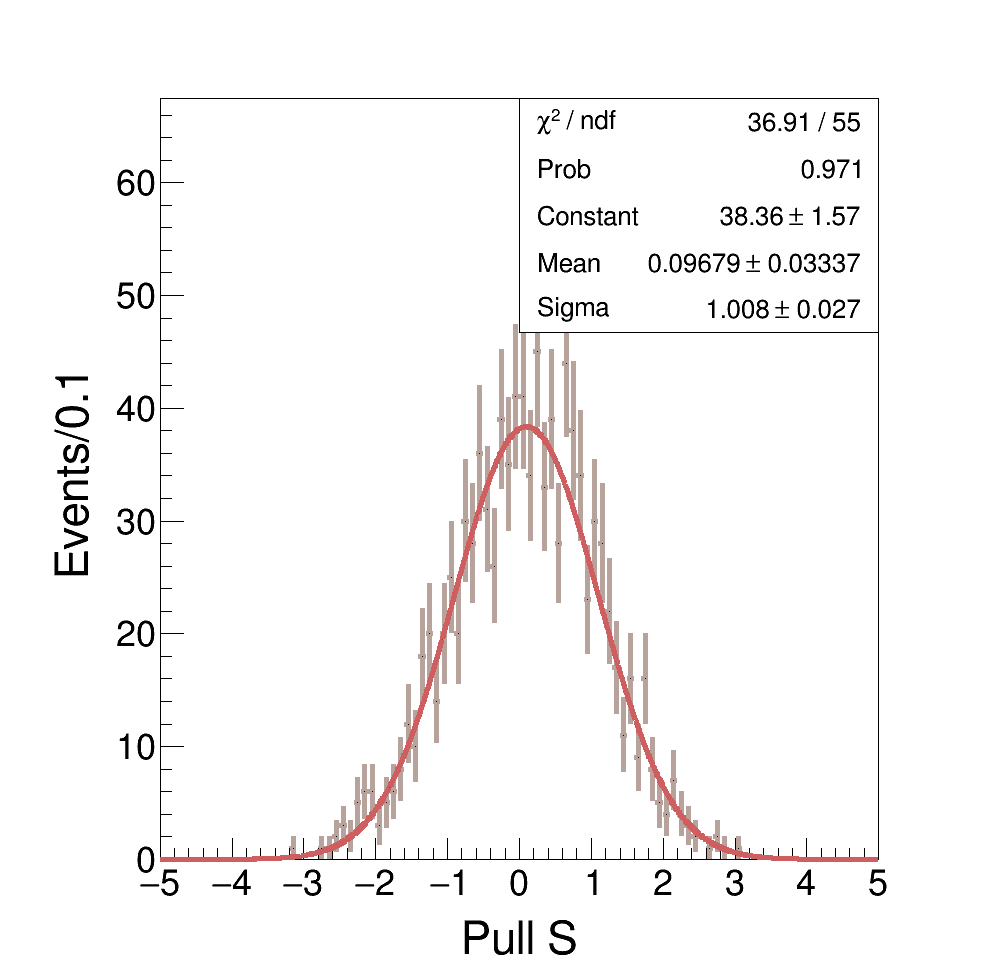}
\caption{Distributions of the fitted value (left) and the pull (right) for \(S_{\phi\gamma}\), obtained from 1000 pseudo-experiments with an input generated value of \(S_{\phi\gamma}^{\mathrm{gen}}=0.43\). The pull is defined as \((S_{\phi\gamma}^{\mathrm{fit}}-S_{\phi\gamma}^{\mathrm{gen}})/\sigma_{S}\), where \(S_{\phi\gamma}^{\mathrm{fit}}\) is the fitted value in each pseudo-experiment, \(S_{\phi\gamma}^{\mathrm{gen}}\) is the generated input value, and \(\sigma_S\) is the fit uncertainty of \(S_{\phi\gamma}^{\mathrm{fit}}\).\label{fig:PullS}}
\end{figure}

The CP-violating parameters \(C_{\phi\gamma}\) and \(S_{\phi\gamma}\) require initial-flavor-tagged samples because of the rapid \(B_s^0\) oscillation and can be extracted from initial-flavor-tagged \(B_s^0\) and \(\bar{B}_s^0\) decays. We combined Eq.~\ref{eq:1} with initial flavor tagging efficiency, acceptance (Eq.~\ref{eq:4}) and time resolution, generated 472K Monte Carlo $B_s^0 \to \phi(\to K^+K^-)\gamma$ events and 68K Monte Carlo background events. The signal events are generated with input values \(A_{\phi\gamma}^{\Delta} = -0.67\), \(C_{\phi\gamma} = 0.11\), and \(S_{\phi\gamma} = 0.43\). The decay time distribution of the combined MC signal and background sample, along with fits, is presented in Figure~\ref{fig:FitCS}. The fit accurately models the complex decay time spectrum and successfully extracts the parameters, yielding $\mathcal{C}_{\phi\gamma} = 0.113 \pm 0.010$ and $\mathcal{S}_{\phi\gamma} = 0.416 \pm 0.010$. We perform 1000 independent pseudo-experiments on statistically independent samples obtained from the same MC distributions. The distribution of fitted \(\mathcal{C}_{\phi\gamma}\) and \(\mathcal{S}_{\phi\gamma}\)  are shown in the left panel of Figure~\ref{fig:PullC} and Figure~\ref{fig:PullS}, respectively. The statistical uncertainty of \(C_{\phi\gamma}\) and \(S_{\phi\gamma}\) are \(\sigma_C^{\text{stat}} = 0.009\) and \(\sigma_S^{\text{stat}} = 0.010\). The pull distribution for the \(\mathcal{C}_{\phi\gamma}\) and \(\mathcal{S}_{\phi\gamma}\) variables are shown in the right panel of Figure~\ref{fig:PullC} and Figure~\ref{fig:PullS}, respectively, the result shown that the bias of fitted C and S are within a reasonable range.

\begin{figure}[h]
\centering
\includegraphics[width=0.48\textwidth]{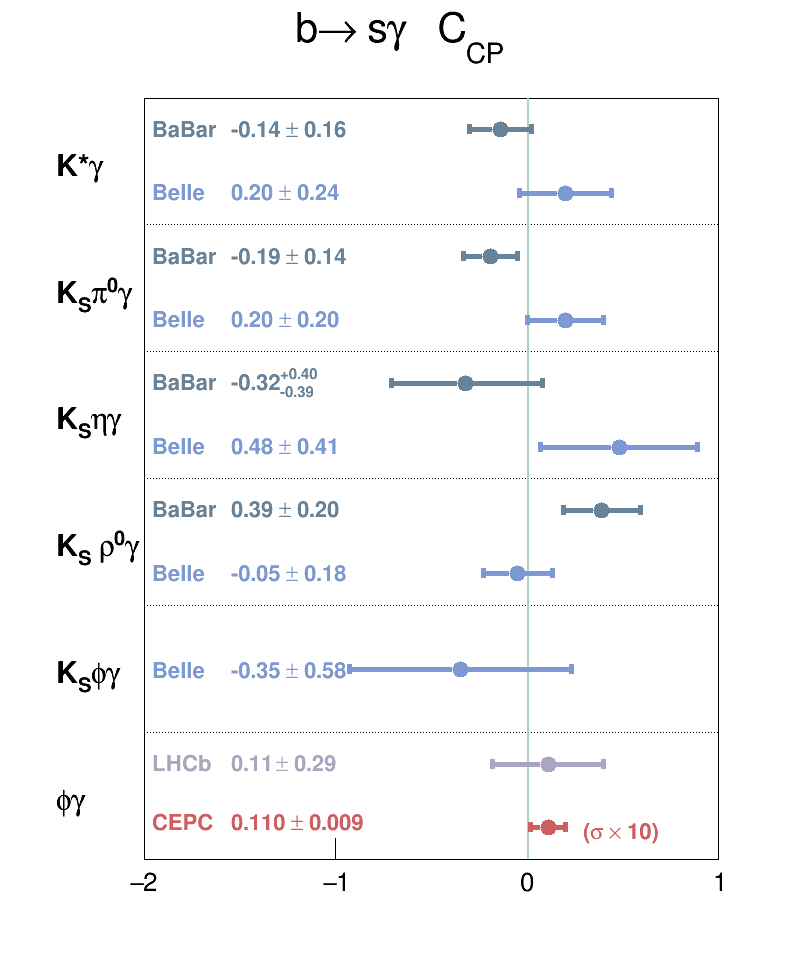}
\includegraphics[width=0.48\textwidth]{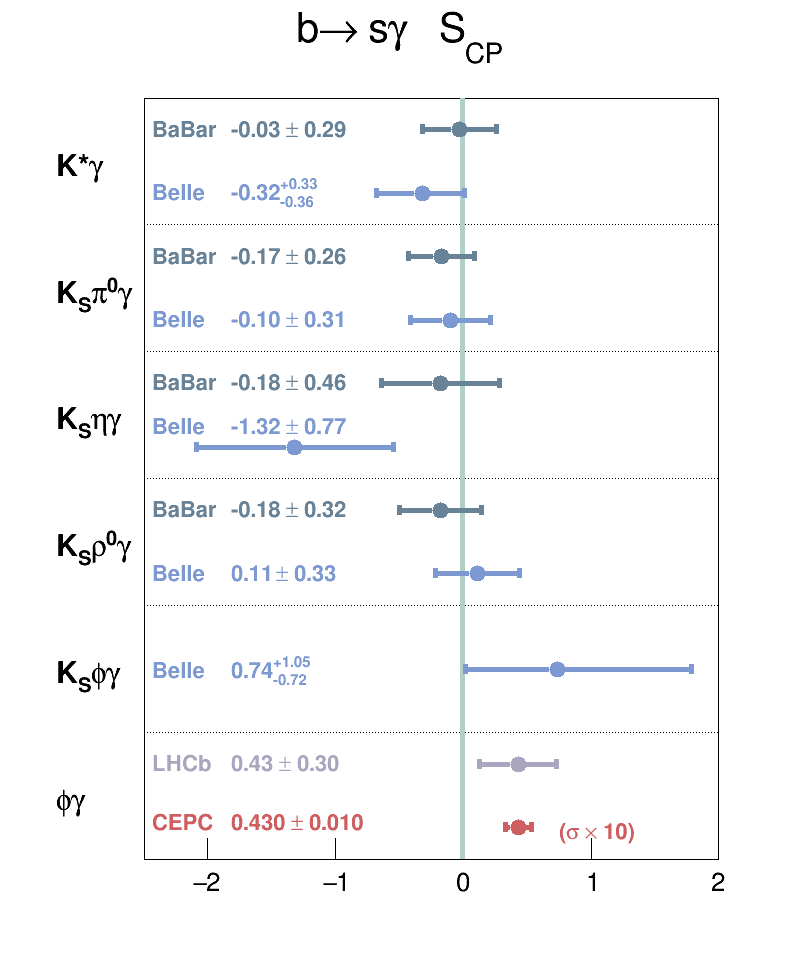}
\caption{Summary of the $C_{CP}$ and $S_{CP}$ statistical uncertainty in $b\to s\gamma$ transitions. To clearly visualize the fine statistical uncertainty of CEPC’s measurements, the error bar for CEPC is magnified by a factor of 10.\label{fig:b2sgamma}}
\end{figure}

Figure \ref{fig:b2sgamma} summarizes the statistical uncertainty of $C_{CP}$ and $S_{CP}$ in $b\to s\gamma$ transitions. The cyan markers indicate the measurement result form BABAR Collaboration \cite{PhysRevD.78.071102,PhysRevD.79.011102,PhysRevD.93.052013}. The blue markers indicate the measurement result form Belle Collaboration \cite{PhysRevD.74.111104,PhysRevD.97.092003,PhysRevLett.101.251601,PhysRevD.84.071101}. The lilac markers indicate the measurement result form LHCb Collaboration \cite{PhysRevLett.123.081802}. The red markers show the CEPC Tera‑Z anticipated uncertainty obtained from this work. To make the small statistical uncertainties visible, the error bars for CEPC are magnified by a factor of 10. The SM predictions are indicated by the green band for comparison. The projected CEPC Tera‑Z results show significantly smaller uncertainties than those from the B‑factories, demonstrating the potential for a significant improvement in precision.

\subsection{Expected systematic uncertainties}
\label{ssec:uncertainty}

To evaluate the expected systematic uncertainties, we adopt a Monte Carlo pseudo-experiment approach, considering three primary sources, which include physical probability density function (PDF) parameters, detector effects and calibration, and background shape modeling. For each systematic source, we generate 1000 independent pseudo-experiments. We perform fits using both the nominal model and the varied model incorporating the specific perturbation. The maximum absolute shift in the mean fitted values of $A_{\phi\gamma}^{\Delta}$, $\mathcal{C}_{\phi\gamma}$, and $\mathcal{S}_{\phi\gamma}$ between the two fit configurations is assigned as the corresponding single-source systematic uncertainty.

For the physical PDF parameters ($\Gamma_s$, $\Delta\Gamma_s$, and $\Delta m_s$), we utilize the anticipated measurement precisions for the CEPC Tera-$Z$ mode. Specifically, the uncertainties for $\Gamma_s$ and $\Delta\Gamma_s$ are set to $\sigma(\Gamma_s) = 0.72 \text{ ns}^{-1}$ and $\sigma(\Delta\Gamma_s) = 2.4 \text{ ns}^{-1}$, respectively, following the detailed prospects studied in Ref.~\cite{Li:2022tlo}. For $\Delta m_s$, the current world average uncertainty from the PDG $\sigma(\Delta m_s) = 5.0 \text{ ns}^{-1}$ is adopted as a conservative input. The systematic shifts are evaluated by varying each parameter by $\pm 1\sigma$ around its central value.

To estimate the impact of detector effects, we consider the decay time acceptance, the time resolution, and the initial-state flavor-tagging. Since the three parameters of the empirical acceptance function ($\alpha$, $\beta$, and $\xi$) exhibit strong non-diagonal correlations, their correlated variations are generated using the Cholesky decomposition of the covariance matrix~\cite{Golub:2013matrix}. The resulting variations are then applied to the acceptance function to evaluate the corresponding systematic uncertainty. The effective time resolution is varied by $\pm 1\sigma$. For the mistag fraction ($\omega$), the LHCb experiment has achieved a relative systematic calibration uncertainty of approximately $0.4\%$ \cite{PhysRevLett.123.081802}. Given the extremely clean di-jet topology at the $e^+e^-$ pole, we conservatively assume that the CEPC will achieve a comparable calibration precision and assign a fixed relative systematic uncertainty of $0.4\%$ to the tagging calibration.

Finally, the uncertainty associated with the background modeling is evaluated by replacing the nominal empirical background PDF with an alternative analytical model. The resulting shifts in the fitted parameters are taken as the systematic uncertainty. Assuming all sources are independent, the total systematic uncertainties are calculated as the sum in quadrature of the individual components, as summarized in Table~\ref{tab:sys_uncertainties}.

\begin{table}[htbp]
  \centering
  \caption{Summary of the expected systematic uncertainties in the analysis of $B_s^0 \to \phi\gamma$ decays at the CEPC.}
  \label{tab:sys_uncertainties}
  \begin{tabular}{llccc}
    \hline\hline
    Category & Source of Uncertainty & $\sigma(A_{\phi\gamma}^{\Delta})$ & $\sigma(\mathcal{C}_{\phi\gamma})$ & $\sigma(\mathcal{S}_{\phi\gamma})$ \\
    \hline
    \multirow{3}{*}{Physics PDF Parameters} 
    & $\Gamma_s$       & $0.0201$ & $0.0002$ & $0.0002$ \\
    & $\Delta\Gamma_s$ & $0.0265$ & $0.0001$ & $0.0004$ \\
    & $\Delta m_s$     & - & $0.0035$ & $0.0007$ \\
    \hline
    \multirow{3}{*}{Detector \& Calibration} 
    & Time acceptance    & $0.0105$ & $0.0002$ & $0.0008$ \\
    & Time resolution    & $0.0027$ & $0.0015$ & $0.0080$ \\
    & Flavor mistag rate (0.4\%) & - & $0.0005$ & $0.0016$ \\
    \hline
    \multirow{1}{*}{Background Modeling} 
    & Shape modeling     & $0.0172$ & $0.0070$ & $0.0054$ \\
    \hline
    \multirow{1}{*}{\textbf{Total}} 
    & $\boldsymbol{\sigma_{\text{syst.}}}$ & $\mathbf{0.0389}$ & $\mathbf{0.0080}$ & $\mathbf{0.0098}$ \\
    \hline\hline
  \end{tabular}
\end{table}

The anticipated results for the CP-violating parameters at the CEPC Tera-$Z$ mode are expressed as:
\begin{align*}
\boldsymbol{A_{\phi\gamma}^\Delta} &= -0.670 \pm 0.021\text{ (stat.)} \pm 0.040\text{ (syst.)}, \\
\boldsymbol{\mathcal{C}_{\phi\gamma}} &= \phantom{-}0.110 \pm 0.009\text{ (stat.)} \pm 0.008\text{ (syst.)}, \\
\boldsymbol{\mathcal{S}_{\phi\gamma}} &= \phantom{-}0.430 \pm 0.010\text{ (stat.)} \pm 0.010\text{ (syst.)}.
\end{align*}

The total uncertainties, obtained by adding the statistical and systematic components in quadrature, are estimated to be $\sigma_{\text{tot}}(A_{\phi\gamma}^{\Delta}) = 0.044$, $\sigma_{\text{tot}}(\mathcal{C}_{\phi\gamma}) = 0.011$, and $\sigma_{\text{tot}}(\mathcal{S}_{\phi\gamma}) = 0.014$. These represent an improvement of approximately one order of magnitude in precision  compared to the current LHCb results \cite{PhysRevLett.123.081802,PhysRevLett.118.021801}.

\subsection{Sensitivity to New Physics}

\begin{figure}[h]
\centering
\includegraphics[width=0.4\textwidth]{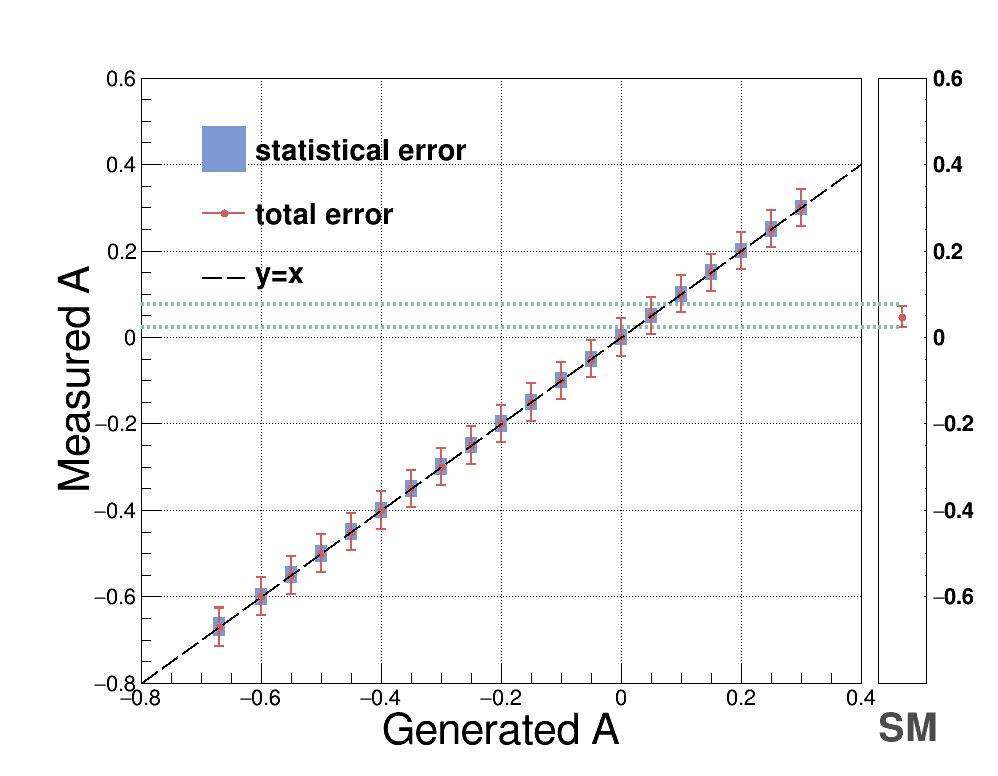}\\
\includegraphics[width=0.4\textwidth]{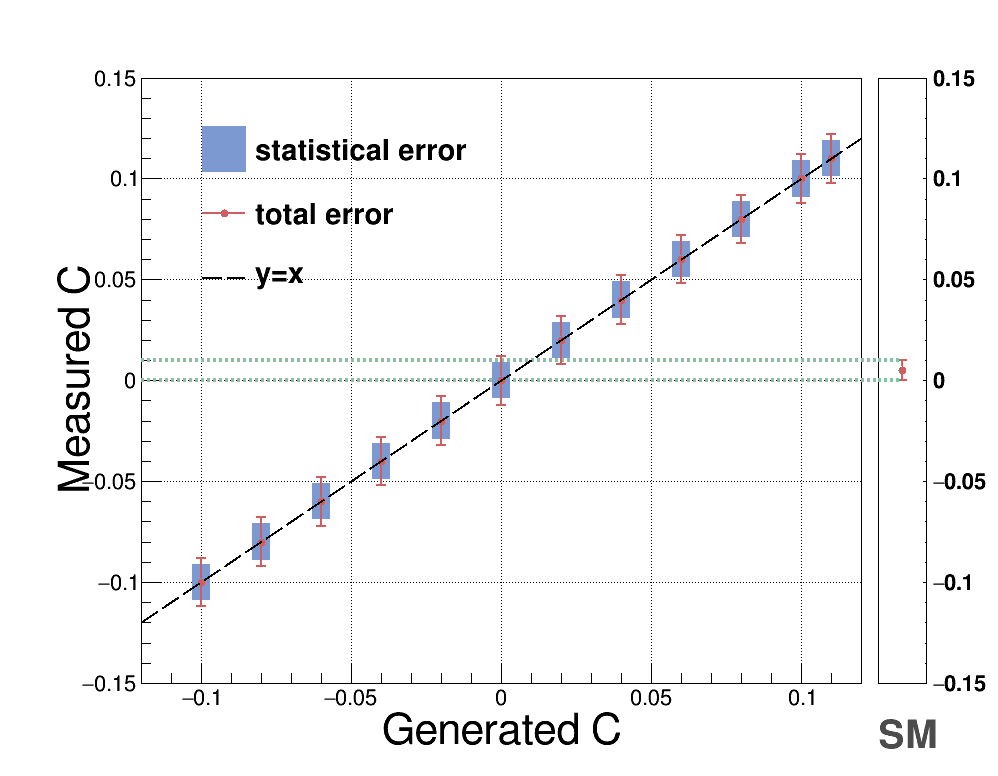}
\includegraphics[width=0.4\textwidth]{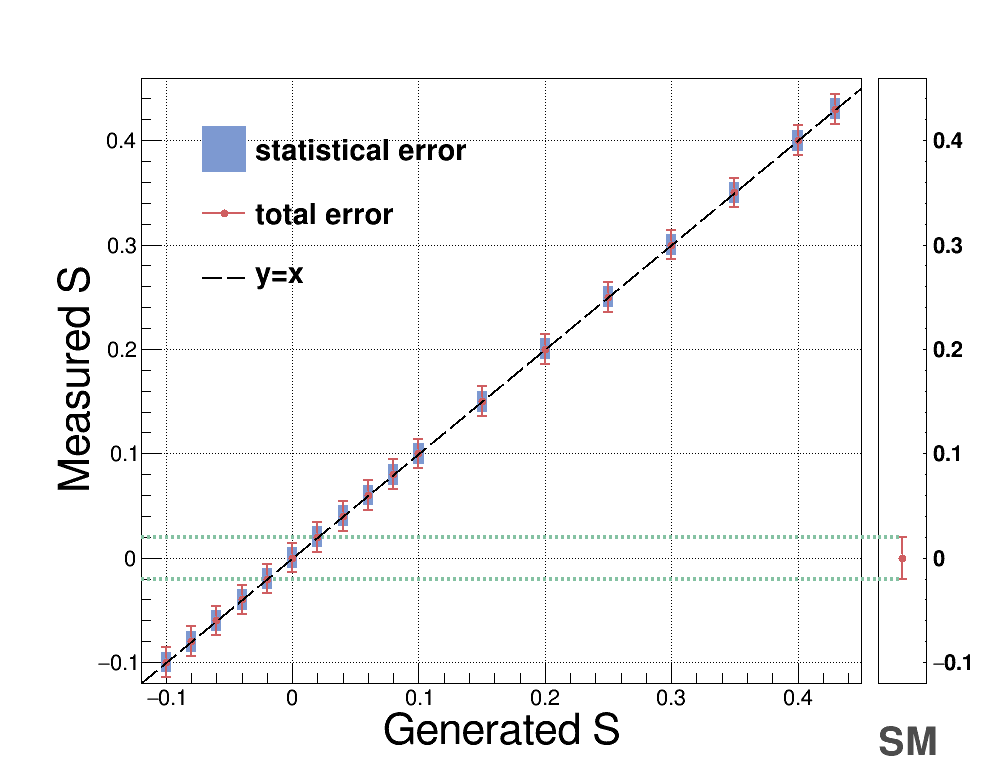}
\caption{Sensitivity scan for the CP-violating parameters \(\mathcal{A}^\Delta\), \(\mathcal{C}\), the blue rectangular areas represent the expected statistical uncertainties, while the red error bars indicate the total expected uncertainties, which combines systematic uncertainty and the statistical uncertainty. The green dashed lines mark the range predicted by the SM. \label{fig:Sensitivity}}
\end{figure}

To investigate the CEPC's sensitivity to potential NP contributions, we performed a parameter scan over \(\mathcal{A}^\Delta\), \(\mathcal{C}\), and \(\mathcal{S}\). The results are illustrated in Figure~\ref{fig:Sensitivity}. When the true parameter values correspond to the SM predictions, the projected total experimental uncertainties for $\mathcal{A}^\Delta$ and $\mathcal{C}$ are slightly larger than the current theoretical uncertainties. However, for $\mathcal{S}$, the projected experimental precision is comparable to the SM theoretical uncertainty.

To clearly establish the criteria for identifying NP, it is essential to define the expected $1\sigma$ confidence intervals assuming the SM holds true. Taking into account the projected total experimental uncertainties of the CEPC and the current theoretical uncertainties, the $1\sigma$ SM-consistent intervals are determined to be approximately: $-0.05 \leq \mathcal{A}_{\phi\gamma}^\Delta \leq 0.15$, $-0.02 \leq \mathcal{C}_{\phi\gamma} \leq 0.04$, and $-0.04 \leq \mathcal{S}_{\phi\gamma} \leq 0.04$.  While establishing evidence for NP naturally requires substantially larger deviations, the significantly reduced uncertainties anticipated at the CEPC dramatically narrows the allowed parameter space. Consequently, any NP models predicting contributions will be severely constrained.

\begin{figure}[h]
\centering
\includegraphics[width=0.45\textwidth]{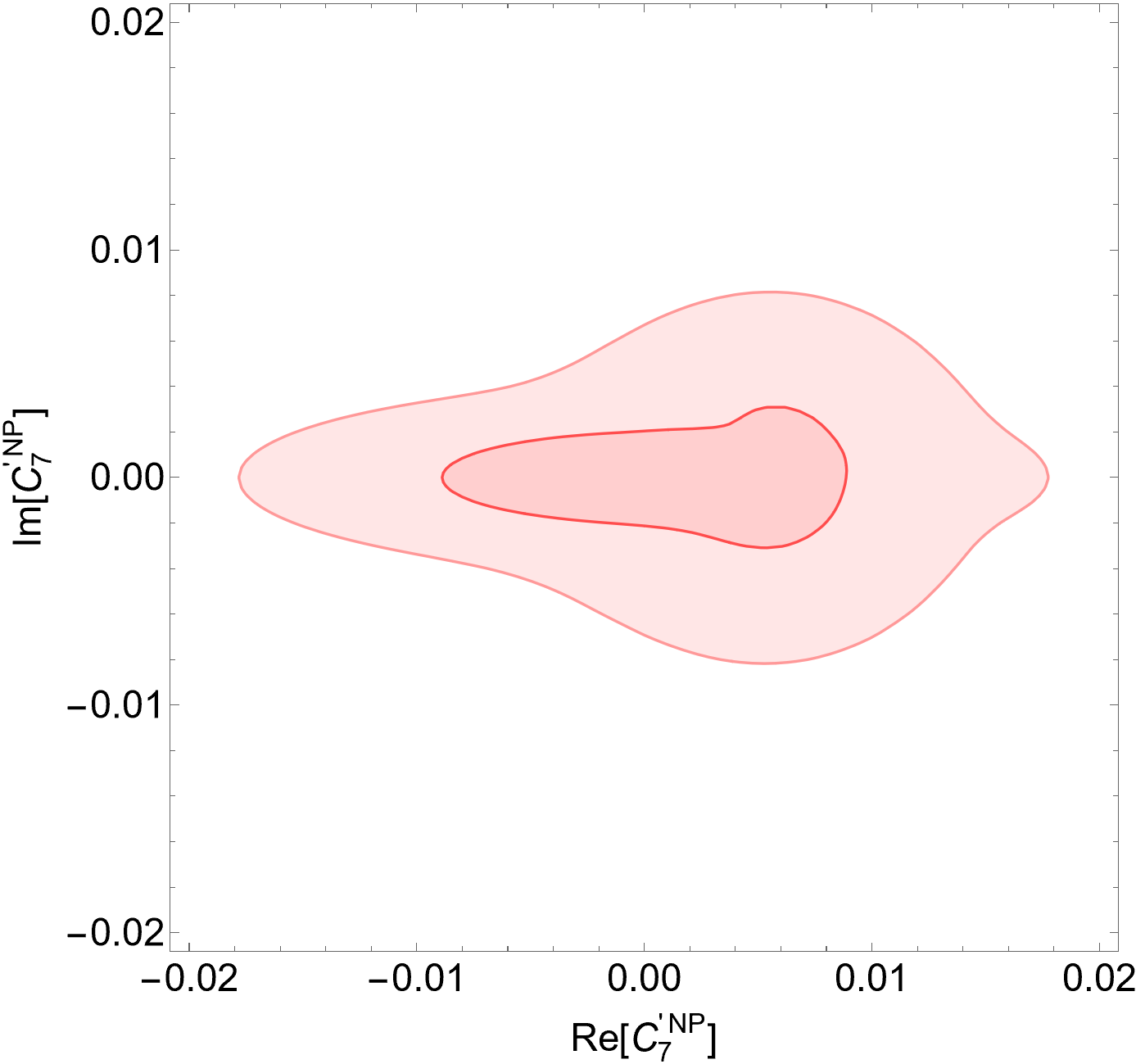}
\includegraphics[width=0.45\textwidth]{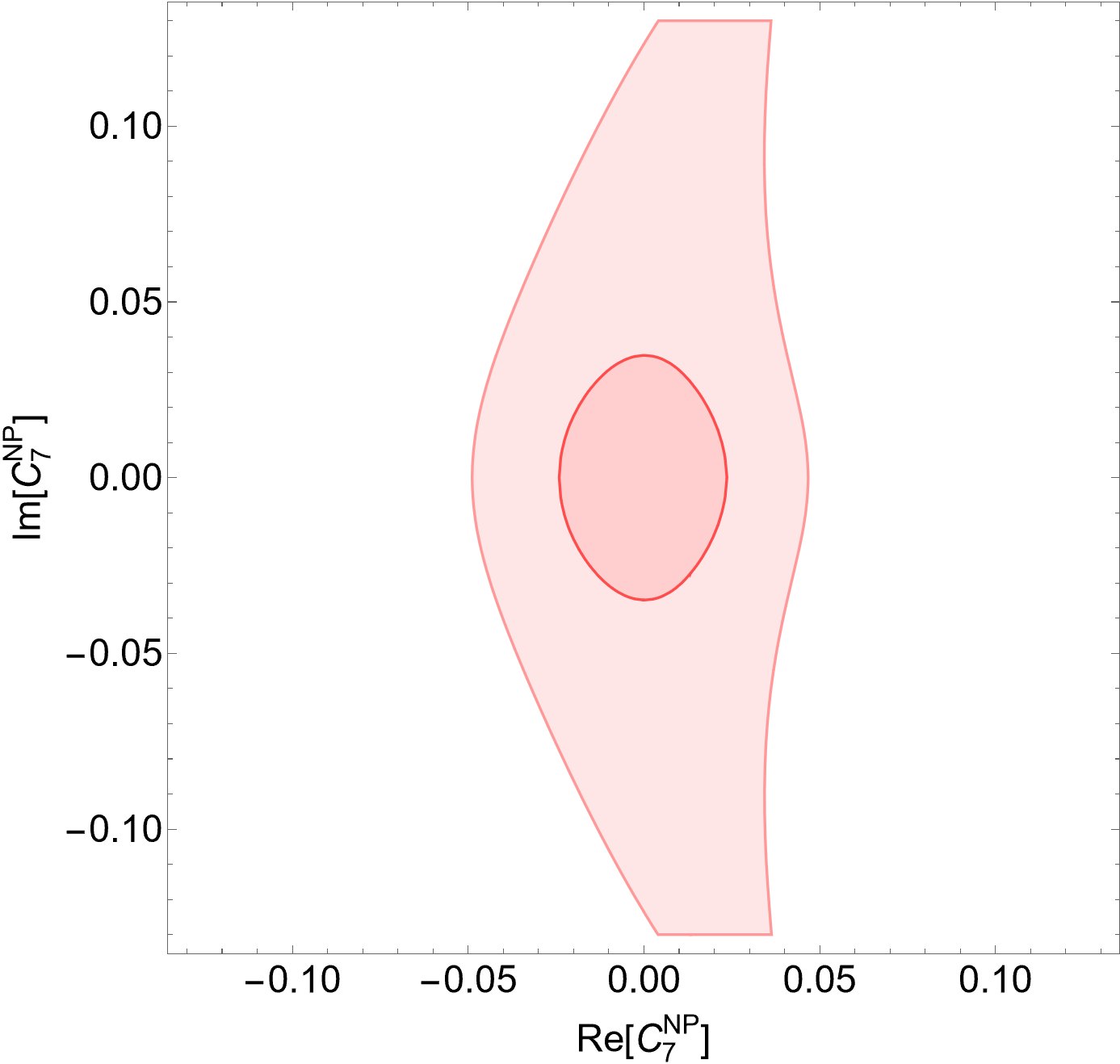}
\caption{Frequentist constraints on New Physics contributions to the complex Wilson coefficients $C_7^{\prime\text{NP}}$ (left) and $C_7^{\text{NP}}$ (right). In both panels, we take the SM as our benchmark and incorporate the overall expected uncertainties derived in Section~\ref{ssec:uncertainty}. Note that the two plots have different axis ranges.\label{fig:Constraint}}
\end{figure}

The measured CP-conserving and CP-violating parameters are useful in constraining the Wilson coefficients of effective field theory (EFT) operators. For example, by comparing the predictions of the SM with the expected experimental uncertainties, the sensitivities of \(B_{s}^{0} \to \phi \gamma\) decays at the CEPC Tera-$Z$ factory can be used to impose constraints on low-energy EFT (LEFT) operators, as well as constraints on UV New Physics contributions~\cite{Altmannshofer:2014rta,Bharucha:2015bzk,Paul:2016urs,Blake:2016olu,Bordone:2025cde}. In this work, we focus on the NP contributions to the dipole operators $\mathcal{O}_7^{(\prime)}$, which are defined in the effective Hamiltonian as:
\begin{equation}
    \mathcal{H}_{\rm eff} \supset  \frac{4 G_F V_{tb}V_{ts}^\ast}{\sqrt{2}}\bigg[C_7 \bigg(\frac{e m_b}{16\pi^2} \bar{s}_L \sigma_{\mu\nu}b_R\bigg)+C_7^\prime \bigg(\frac{e m_b}{16\pi^2} \bar{s}_R \sigma_{\mu\nu}b_L\bigg) \bigg] F^{\mu\nu} + \dots \label{eq:Heff}
\end{equation}
where $G_F$ is the Fermi constant, $V_{tb}$ and $V_{ts}^\ast$ are the relevant Cabibbo-Kobayashi-Maskawa (CKM) matrix elements, $e$ is the electromagnetic coupling constant, $m_b$ is the bottom quark mass, $s_{L/R}$ and $b_{R/L}$ denote the left- and right-handed quark fields, $\sigma_{\mu\nu}$ is the Dirac tensor, and $F^{\mu\nu}$ is the electromagnetic field tensor. The omitted terms represent other flavor-changing neutral current (FCNC) operators.

In the SM, $C_7$ is generated at the leading loop order, while $C_7^\prime$ is helicity suppressed and therefore smaller by a factor of $\mathcal{O}(m_s/m_b)$. At higher loop orders, four-fermion operators also contribute to the effective value of $C_7$ ($C_7^{\rm eff}$)~\cite{Chetyrkin:1996vx}, leaving considerable theoretical uncertainties in the calculation of $C_7$. However, the time-dependent CP-violating observables $\mathcal{C}$ and $\mathcal{S}$ are less sensitive to these (soft) hadronic uncertainties, allowing for a cleaner and more precise determination of the ratios $\text{Re}(C_7^\prime/C_7)$ and $\text{Im}(C_7^\prime/C_7)$. Since performing a global fit involving other $b\to s$ transition measurements is beyond the scope of this work, we adopt a simplified approach: we fit the NP contributions to $C_7$ and $C_7^\prime$ separately. Specifically, when fitting the NP contribution for one operator, the NP contribution to the other is strictly set to zero. For the theoretical values of the three CPV observables and their corresponding theoretical uncertainties, we follow the calculations provided in Refs.~\cite{Artuso:2015swg,Paul:2016urs}.

The fit results are shown in Figure~\ref{fig:Constraint}, presenting the projected frequentist constraints on the NP contributions given the anticipated CEPC Tera-$Z$ sensitivity. To obtain these contours, for each point in the 2D complex plane, we evaluate the minimum $\chi^2$ value by assuming that the central values of all four measured observables (Branching Fraction, $\mathcal{A}^\Delta$, $\mathcal{C}$, $\mathcal{S}$) align exactly with their corresponding SM predictions. This calculation incorporates both the expected statistical uncertainties and the overall systematic uncertainties (combining theoretical and experimental sources). The left and right panels display the constraints on $C_7^{\prime\text{NP}}$ and $C_7^{\text{NP}}$, respectively. The lighter and darker shaded regions correspond to the $1\sigma$ and $2\sigma$ allowed confidence intervals. Notably, the CP-violating observables exhibit exceptional sensitivity to NP contributions associated with $C_7^{\prime\text{NP}}$. It is worth emphasizing that the constraints presented herein are derived solely from the \(B_{s}^{0} \to \phi \gamma\) channel at the CEPC Tera-$Z$; complementary and potentially stronger constraints from other channels, such as \(B_{s}^{0} \to \phi \ell^+\ell^-\), are expected in future global analyses~\cite{Kwok:2025fza}.

\section{Conclusion}
 
The \(b \to s\gamma\) transition, being a typical FCNC process, serves as a crucial probe for detecting CP violation and NP. Especially in the context of future $Z$ factories, its study is highly important for probing new physics beyond the SM. The CEPC, leveraging its high statistics, clean collision environment, and excellent detector performance, offers unique advantages for flavor physics research and is an ideal platform for investigating \(b \to s\gamma\) mediated decays.  

This study focuses on the FCNC decay \(B_{s}^{0} \to \phi \gamma\) and presents a feasibility study under the CEPC Tera-$Z$ operation scenario. The results demonstrate that the CEPC can achieve an anticipated relative statistical uncertainty of approximately $(0.107 \pm 0.004)\%$. While the ultimate measurement at this precision level will inevitably be constrained by systematic uncertainties, this unprecedented statistical reach represents an improvement of roughly two orders of magnitude compared to existing measurements. The corresponding signal selection efficiency and purity are $45\%$ and $93\%$, respectively, with an invariant mass resolution of $98.0$~$\mathrm{MeV}/c^2$. This precision level lays a solid foundation for the subsequent extraction of CP-violating parameters and sensitive searches for NP effects. 

Accounting for the $B^{0} $-$\bar{B}^{0}$ mixing effect, a time-dependent analysis of the \(B_s^0 \to \phi\gamma\) decay was performed, and the mixing-induced CP-violating parameters \(\mathcal{A}_{\phi\gamma}^\Delta\), \(\mathcal{C}_{\phi\gamma}\), and \(\mathcal{S}_{\phi\gamma}\) were extracted. Using the central values of the LHCb measurements as inputs, the anticipated precisions of these parameters are evaluated as:
\begin{align*}
\boldsymbol{A_{\phi\gamma}^\Delta} &= -0.670 \pm 0.021\text{ (stat.)} \pm 0.040\text{ (syst.)}, \\
\boldsymbol{\mathcal{C}_{\phi\gamma}} &= \phantom{-}0.110 \pm 0.010\text{ (stat.)} \pm 0.008\text{ (syst.)}, \\
\boldsymbol{\mathcal{S}_{\phi\gamma}} &= \phantom{-}0.430 \pm 0.010\text{ (stat.)} \pm 0.010\text{ (syst.)}.
\end{align*}
These anticipated precisions provide an opportunity for testing CP-violating mechanisms within the SM and searching for NP contributions. Furthermore, based on the total expected uncertainties, the $1\sigma$ SM-consistent intervals are determined to be approximately $-0.05 \leq \mathcal{A}_{\phi\gamma}^\Delta \leq 0.15$, $-0.02 \leq \mathcal{C}_{\phi\gamma} \leq 0.04$, and $-0.04 \leq \mathcal{S}_{\phi\gamma} \leq 0.04$. Any NP models predicting right-handed helicity contributions that exceed these stringent boundaries will be severely constrained.  

In summary, this study systematically demonstrates the CEPC's advantages in the research of the \(B_s^0 \to \phi\gamma\) decay and related flavor physics, such as FCNC processes and rare decays. The projected high precision for both signal reconstruction and CP-violating parameter extraction establishes stringent constraints on the SM, which inherently serves as a highly sensitive window for probing NP effects. Nevertheless, it is important to acknowledge that such projected precision is subject to certain phenomenological conditions. This paper adopts the fast simulation method and assumes a parameterized detector response, while realistic full-simulation detector effects will introduce additional impacts. Background modeling depends on the hadronization models in the generators, and the precise calibration accuracy of jet flavor tagging will further affect the ultimate systematic precision. These aspects merit comprehensive investigations in future full-simulation studies.

% Bibliography

\section*{Acknowledgements}
This work was supported by National Key Program for S\&T Research and Development under contract number 2024YFA1610603 and 2022YFA1601901. We thank the HEPAI platform of the Institute of High Energy Physics for providing the computing resources. We also thank Wolfgang Altmannshofer, Yuzhi Che and Chia-Wei Liu for useful discussions. 

%% [A] Recommended: using JHEP.bst file
\bibliographystyle{JHEP}
\bibliography{biblio.bib}

%% or
%% [B] Manual formatting (see below)
%% (i) We suggest to always provide author, title and journal data or doi:
%% in short all the informations that clearly identify a document.
%% (ii) please avoid comments such as "For a review'', "For some examples",
%% "and references therein" or move them in the text. In general, please leave only references in the bibliography and move all
%% accessory text in footnotes.
%% (iii) Also, please have only one work for each \bibitem.

% \begin{thebibliography}{99}

% \bibitem{a}
% Author,
% \emph{Title},
% \emph{J. Abbrev.} {\bf vol} (year) pg.

% \bibitem{b}
% Author,
% \emph{Title},
% arxiv:1234.5678.

% \bibitem{c}
% Author,
% \emph{Title},
% Publisher (year).

% \end{thebibliography}
\end{document}